\documentstyle[12pt]{article}
\def\ifundefined#1{\expandafter\ifx\csname#1\endcsname\relax}
\makeatletter \ifundefined{new@mathgroup} {} \else   %\input{oldlfont.sty}\fi
    \new@mathgroup\sffam
    \new@internalmathalphabet\mathsf\sffam{cmss}{m}{n}
    \def\psf{\fontfamily\sfdefault \fontseries\default@series
        \fontshape\default@shape\selectfont\mathsf}
\fi

\def\citen#1{\if@filesw \immediate\write \@auxout {\string\citation{#1}}\fi%
\@tempcntb\m@ne \let\@h@ld\relax \def\@citea{}%
\@for \@citeb:=#1\do {\@ifundefined {b@\@citeb}%
    {\@h@ld\@citea\@tempcntb\m@ne{\bf ?}%
    \@warning {Citation `\@citeb ' on page \thepage \space undefined}}%
    {\@tempcnta\@tempcntb \advance\@tempcnta\@ne
    \setbox\z@\hbox\bgroup\ifcat0\csname b@\@citeb \endcsname \relax
    \egroup \@tempcntb\number\csname b@\@citeb \endcsname \relax
    \else \egroup \@tempcntb\m@ne \fi \ifnum\@tempcnta=\@tempcntb
    \ifx\@h@ld\relax \edef \@h@ld{\@citea\csname b@\@citeb\endcsname}%
    \else \edef\@h@ld{\hbox{--}\penalty\@highpenalty
    \csname b@\@citeb\endcsname}\fi
    \else \@h@ld\@citea\csname b@\@citeb \endcsname \let\@h@ld\relax \fi}%
\def\@citea{,\penalty\@highpenalty\hskip.13em plus.13em minus.13em}}\@h@ld}
\def\@citex[#1]#2{\@cite{\citen{#2}}{#1}}%
\def\@cite#1#2{\leavevmode\unskip\ifnum\lastpenalty=\z@\penalty\@highpenalty\fi%
  \ [{\multiply\@highpenalty 3 #1%
  \if@tempswa,\penalty\@highpenalty\ #2\fi}]}   %
\makeatother % end of CITE.STY

\def\alg           {algebra}
\newcommand{\andauthoretc}[5]{\\[2 mm]{}\centerline{and}{}\\[2 mm]
                   \centerline{\sc #1}\\[2 mm]
                   \centerline{#2}\\[.5 mm] \centerline{#3}\\[.5 mm]
                   \centerline{#4}\\[.5 mm] \centerline{#5}}
\long\def\authoretc#1#2#3#4#5{\centerline{\sc #1}\\[2 mm]{}%
                   \centerline{#2}\\[.5 mm] \centerline{#3}\\[.5 mm]
                   \centerline{#4}\\[.5 mm] \centerline{#5}}
\newcommand{\arrayleft}[1]{$\left\{\begin{array}{l} #1 \end{array}\right.$}

\def\be            {\begin{equation}}
\newcommand{\bin}[2]{\left(\begin{array}{c} #1\\ #2 \end{array}\right)}
\newcommand{\biN}[2]{\Llb\!\!\begin{array}{c} {\scriptstyle #1}\\[-1.3 mm]
                   {\scriptstyle #2} \end{array}\!\!\Lrb}
\def\bfe           {{\bf1}}

\def\bke           {\mbox{$(B_k)_{2n}$}}
\def\bkn           {\mbox{$(B_k)_{2n+1}$}}
\def\Bkn           {\mbox{$(B,2k+1,2n+1)$}}

\def\bnk           {\mbox{$(B_n)_{2k+1}$}}
\def\Bnk           {\mbox{$(B,2n+1,2k+1)$}}

\def\cdim          {conformal dimension}
\def\cft           {conformal field theory}
\def\cfts          {conformal field theories}
\def\Chi           {{\cal X}}
\def\ckn           {\mbox{$(C_k)_n$}}
\def\Ckn           {\mbox{$(CC,k+1,n-1)$}}
\def\ckri          {\mbox{$\breve\chi^{}_f$}}
\def\cl            {\mbox{$c^{}_\Lambda$}}
\def\clp           {\mbox{$c^{}_{\Lambda'}$}}
\let\cli=\centerline
\def\cnk           {\mbox{$(C_n)_k$}}
\def\Cnk           {\mbox{$(CC,n,k)$}}
\long\def\coset#1#2#3{\mbox{${\cal C}[#1/ #2]_{#3}^{}$}}

\long\def\cosetK#1#2{\mbox{${\cal C}[#1/ #2]_K^{}$}}

\newcommand{\CSfull}[5]{${\cal C}[ (#1)_K\oplus (D_{#2})_1 \,/\,
                   (#3)_{#4} \oplus (u_1)_{#5}]$}

\newcommand{\CSfulld}[7]{${\cal C}[(#1)_K\oplus (D_{#2})_1 \,/\,
                   (#3)_{#4} \oplus(#5)_{#6} \oplus (u_1)_{#7}]$}
\def\csp           {\,/\,}

\def\ct            {coset theory}
\def\cts           {coset theories}

\def\detij         {\mbox{${\rm det}_{i,j}$}}

\newcommand{\dfrac}[2]{{\displaystyle\frac{#1}{#2}}}

\long\def\del#1    \enddel{}

\def\dhlind#1      {\\[-#1 mm]&&&\\ \hline\hline  {}&&&\\[-#1 mm] }
\def\dhline#1      {\\[-#1 mm]&\\ \hline\hline  {}&\\[-#1 mm] }
\def\dhlinf#1      {\\[-#1 mm]&&&&&\\ \hline\hline  {}&&&&&\\[-#1 mm] }
\def\dhlinv#1      {\\[-#1 mm]&&&&\\ \hline\hline  {}&&&&\\[-#1 mm] }
\def\dhlinz#1      {\\[-#1 mm]&&\\ \hline\hline  {}&&\\[-#1 mm] }

\def\dnk           {\mbox{$(D_n)_{2k+1}$}}

\def\ee            {\end{equation}}
\def\eE            {{\rm e}}
\def\ele           {\mbox{$\ell_1$}}
\def\eli           {\mbox{$\ell_i$}}
\def\elI           {\mbox{$\ell^{(\tau)}_i$}}
\let\emb=\hookrightarrow

\def\epop          {extended Poin\-car\'e polynomial}
\let\eps=\varepsilon
\newcommand{\erf}[1]{(\ref{#1})}
\def\fl            {f(\Lambda')}
\newcommand{\fline}[1]{\vfill\noindent ------------------\\[1 mm]}
\def\furu          {fusion rule}
\def\futnote#1     {\footnote{~#1}\ }

\def\gv            {\mbox{$g_{}^\vee$}}

\def\heq           {\,\hat=\,}
\def\hing          {\mbox{$h\emb g$}}

\def\hl            {\mbox{$\Delta_\Lambda$}}
\def\htl           {\mbox{$\Delta_{\tau(\Lambda)}$}}

\def\hsc           {hermitian symmetric coset}
\newcommand{\hsp}[1] {\mbox{\hspace{#1 mm}}}
\def\hy            {$\mbox{-\hspace{-.66 mm}-}$}
\def\Ibar          {\overline I}

\def\ii            {{\rm i}}
\def\Jbar          {\overline J}
\def\jc            {\mbox{$J_{\rm c}$}}
\def\jf            {J.\ Fuchs}
\def\js            {\mbox{$J_{\rm s}$}}
\def\jv            {\mbox{$J_{\rm v}$}}

\def\kpf           {Kac\hy Peterson formula}

\def\L             {\mbox{$\Lambda$}}
\newcommand{\la}[1]{\mbox{$\Lambda_{(#1)}$}}

\def\clearpag{\typeout{}\typeout{}\typeout{here clearpage - if your TeX memory}
\typeout{is large enough, you can remove this}\typeout{}\typeout{} \clearpage}

\def\latw          {\mbox{$\tilde{\lambda}$}}
\def\li            {\mbox{$\Lambda_{(i)}$}}
\def\lie           {Lie algebra}

\def\llb           {\mbox{\large(}}
\def\lrb           {\mbox{\large)}}
\def\Llb           {\mbox{\Large(}}
\def\Lrb           {\mbox{\Large)}}
\def\lto           {\mbox{$\tilde\ell_\circ$}}

\def\mand          {\mbox{and}}
\def\mfor          {\mbox{for}}

\def\mijll         {\mbox{${\cal M}_{ij}(\Lambda,\Lambda')$}}
\def\mijmll        {\mbox{${\cal M}^-_{ij}(\Lambda,\Lambda')$}}
\def\mijpll        {\mbox{${\cal M}^+_{ij}(\Lambda,\Lambda')$}}
\def\mijtll        {\mbox{$\tilde{\cal M}_{ij}(\Lambda,\Lambda')$}}
\def\ml            {\mbox{$M_\Lambda$}}
\def\mL            {M_\Lambda}

\def\mtl           {\mbox{$M_{\tau(\Lambda)}$}}
\def\mtL           {M_{\tau(\Lambda)}}

\def\nijk          {\mbox{${\cal N}_{ij}^{\ \;k}$}}
\def\nn            {$N=2$\ }

\def\NS            {Neveu\hy Schwarz\ }
\def\Om            {\mbox{$A$}}
\newcommand{\omij}[2] {\mbox{$A_{ij}^{({\rm#1#2})}$}}
\def\one           {\mbox{\small $1\!\!$}1}
\def\onehalf       {\mbox{$\frac12$}}
\let\op=\oplus
\def\opa           {operator product algebra}
\def\opop          {ordinary Poin\-car\'e polynomial}

\def\pop           {Poin\-car\'e polynomial}

\def\qtau          {\mbox{$Q_{\Tau}$}}
\long\def\query#1{\hskip 0pt{\vadjust{\everypar={}\small\vtop to 0pt{\hbox{}%
     \vskip -13pt\rlap{\hbox to 46.3pc{\hfil{\vtop{\hsize=8pc\tolerance=6000%
     \hfuzz=.5pc\rightskip=0pt plus 3em\noindent#1}}}}\vss}}}}%

\long\def\rank#1   {\mbox{rank}\,#1}

\def\rep           {representation}

\def\rhog          {\mbox{$\rho_g$}}
\def\rhoh          {\mbox{$\rho_h^{}$}}
\def\rhot          {\mbox{$\rho^{(\tau)}$}}
\def\rgs           {Ramond ground state}
\newcommand{\sect}[1] {\section{#1}\setcounter{equation}{0}}
\def\sign          {\mbox{sign\,}}

\def\sll           {\mbox{$S^{}_{\Lambda,\Lambda'}$}}
\def\sllp          {\mbox{$S^{\scriptscriptstyle(+)}_{\Lambda,\Lambda'}$}}
\def\stll          {\mbox{$S^{}_{\tau(\Lambda),\tau(\Lambda')}$}}
\def\smat          {$S$-matrix}

\def\sps           {spinor-symmetric}
\def\spS           {{\rm s}}

\def\sumin         {\sum_{i=1}^n}

\def\sumW          {\sum_{w\in W}}

\def\Tau           {{\cal T}}
\def\tgll          {\mbox{$\tau(\Lambda,\lambda)_>$}}
\def\tglp          {\mbox{$\tau(\Lambda',\lambda')_>$}}

\newcommand{\titleetC}[6] { {\tt #1}\mbox{}\\ \rightline{{\sf\prepnr}}
     \\ \rightline{{\sf\prepnR}} {}\\ \rightline{{\sf hep-th/9307107}} {}\\
     \rightline{{\sf #2}} {}\\[16 mm]%
     \cli{\bf\Large {#3}}\\[2.3 mm] \cli{\bf\Large {#4}}\\[11 mm]{}%
     {#5} \\[18 mm]{}%
     \begin{quote}{\bf Abstract.}\ \,{#6}\end{quote} \newpage}%
\def\tkll          {\mbox{$\tau(\Lambda,\lambda)_<$}}
\def\tklp          {\mbox{$\tau(\Lambda',\lambda')_<$}}
\def\tl            {\mbox{$\tau(\Lambda)$}}
\def\tle           {\mbox{$\tilde\ell_1$}}
\def\Tle           {\mbox{$\tilde\ell_1^{(\tau)}$}}
\def\tli           {\mbox{$\tilde\ell_i$}}
\def\Tli           {\mbox{$\tilde\ell_i^{(\tau)}$}}
\def\twodim        {two-di\-men\-si\-o\-nal}

\def\uE            {\mbox{u$_1$}}

\def\wrt           {with respect to\ }
\def\wtau          {\mbox{$w^{}_{\Tau}$}}
\def\wzw           {WZW\ }
\def\WZW           {Wess\hy Zumino\hy Witten}
\def\wzwm          {WZW model}
\def\wzwt          {WZW theory}
\def\wzwts         {WZW theories}
\def\xns           {\mbox{${\rm x}\in\{0,{\rm v}\}$}}
\def\xr            {\mbox{${\rm x}\in\{{\rm s,c}\}$}}
\def\xtau          {\mbox{${\rm x}_{\Tau}$}}

\def\zet           {{\bf Z}}

   \newcommand{\wb}       {\,\linebreak[0]}
   
   \newcommand{\J}[1]     {{{#1}}\vyp}
   
   \newcommand{\Bi}[1]    {\bibitem{#1}}
   \newcommand{\Prep}[2]  {preprint {#1}}

   \newcommand{\BOOK}[4]  {{\em #1\/} ({#2}, {#3} {#4})}
   \newcommand{\inBO}[7]  {in:\ {\em #1}, {#2}\ ({#3}, {#4} {#5}), p.\ {#6}}
   
   \newcommand{\vyp}[4]   {\ {#1} ({#2}) {#3}}
   \newcommand{\vypf}[5]  {\ {#1} [FS{#2}] ({#3}) {#4}}
   \newcommand{\adma} {Adv.\wb Math.}
   \newcommand{\comp} {Com\-mun.\wb Math.\wb Phys.}

   \newcommand{\ijmp} {Int.\wb J.\wb Mod.\wb Phys.\ A}
   \newcommand{\jopa} {J.\wb Phys.\ A}
   \newcommand{\jomp} {J.\wb Math.\wb Phys.}

   \newcommand{\mqft}[2] {\inBO{Modern Quantum Field Theory} {S.\ Das
              et al, eds.} \WS\Si{1991} {{#1}}{{#2}} }
   \newcommand{\npbf} {Nucl.\wb Phys.\ B\vypf}
   
   \newcommand{\nupb} {Nucl.\wb Phys.\ B}
   \newcommand{\phlb} {Phys.\wb Lett.\ B}
   
   \newcommand{\rmap} {Rev.\wb Math.\wb Phys.}
   \def\CUP    {{Cambridge University Press}}
   \def\WS     {{World Scientific}}
   \def\Ca     {{Cambridge}}
   \def\Si     {{Singapore}}
\def\FBZ {\framebox(19.02,19.02){}}
\newcommand{\bP}[4]  {\begin{picture}(#1,#2)(#3,#4)}
\newcommand{\eP}     {\end{picture}}

\newcommand{\fdx}[2] {\put(#1,#2){\dashbox{2}(20,20){}}}
\newcommand{\fbX}[2] {\thicklines\put(#1,#2){\FBZ}\thinlines}

\newcommand{\hdx}[3] {\multiput(#1,#2)(20,0){#3}{\dashbox{2}(20,20){}}}
\newcommand{\hbX}[3] {\thicklines\multiput(#1,#2)(20,0){#3}{\FBZ}\thinlines}
\newcommand{\mbx}[3] {\put(#1,#2){\makebox(0,0){#3}}}

\newcommand{\vbX}[3] {\thicklines\multiput(#1,#2)(0,20){#3}{\FBZ}\thinlines}
\newcommand{\vdx}[3] {\multiput(#1,#2)(0,20){#3}{\dashbox{2}(20,20){}}}

\newcommand{\labl}[1]{\label{#1}\ee}

\def\prepnr{NIKHEF-H/93-16a} \def\prepnR{HD-THEP-93-27a}

\begin{document}

\titleetC %{}
 { \mbox{}% c.s.\ 15.7.93 \ \ j.f.\ 13.7.93 \ \ DRAFT %-- NOT FOR DISTRIBUTION
  \\[-23 mm] {} }
 {July 1993} {LEVEL-RANK DUALITY OF WZW THEORIES AND} {ISOMORPHISMS
 OF \nn\ COSET MODELS}
 {\authoretc{\ J\"urgen Fuchs\ $^1$}{NIKHEF-H}{Kruislaan 409}
    {NL -- 1098 SJ~~Amsterdam}{}
    \andauthoretc{\ Christoph Schweigert\ $^2$}{Institut f\"ur
       Theoretische Physik} {der Universit\"at Heidelberg}
       {Philosophenweg 16} {D -- 69120~~Heidelberg}}
 {Mappings between certain infinite series of \nn
 superconformal coset models are constructed. They make use of
 level-rank dualities for $B$, $C$ and $D$ type WZW theories,
 which are described in some detail.
 The WZW level-rank dualities do not constitute isomorphisms of the
 theories; for example, for $B$ and $D$ type \wzwts, only simple current
 orbits rather than individual primary fields are mapped onto each other.
 Nevertheless they lead to level-rank dualities of \nn coset
 models that preserve the \cft\ properties in such a manner that the
 coset models related by duality are expected to be in fact
 isomorphic as conformal field theories; in particular,
 the representation of the modular group on the characters and the
 ground states of the Ramond sector are shown to coincide.
 The construction also gives some further insight in the nature of
 the resolution of field identification fixed points of coset theories.
{}\fline{} $^1$~~Heisenberg fellow\\[.2 mm]
$^2$~~supported by Studienstiftung des deutschen Volkes}
\addtolength{\baselineskip}{.45 mm}

\sect{Introduction and summary}

Level-rank dualities relate objects that are present in two different
structures that are connected to each other by exchanging the level
(or possibly some simple function thereof) and the rank of an
affine Lie \alg\ (or some closely related algebraic structure). They
emerge in various areas of physics and mathematics: in \wzw \cfts
\cite{nasc,nasc2,vers2,mnrs,nats} and the theories obtained from them via the
coset construction \cite{albs}; in three-dimensional Chern\hy Simons
theories \cite{mnrs,nars2}; in the \rep\ theory
of quantum groups with deformation parameter a root of unity
\cite{fuva,saal} and of Hecke \alg s whose parameter is a root of unity
\cite{gowe}; and
in the description of edge variables in fusion-RSOS models \cite{kuna}.

Usually, level-rank duality merely implies certain non-trivial relations
among quantities of different theories, such as correlation functions
or \furu s of \wzwm s. In this paper, we describe several level-rank
dualities which go much beyond such relations
in that they provide an isomorphism between
the respective theories. We show that there exist several such
equivalences among infinite
series of \nn superconformal \cts. More specifically, we describe
the identifications
  \be  \begin{array}{l} \Bnk \, \cong \, \Bkn \,, \\[1.5 mm]
  (B,2n, 2k+1) \, \cong \, (B, 2k+1, 2n)_{\mid D} \,, \\[1.5 mm]
  (BB, n+2, 1) \cong (CC, 2, 2n+1) \,,   \\[1.5 mm]
  \Cnk \, \cong \, \Ckn \,.   \\[1.5 mm]
  \end{array} \labl1
Here the notations are taken from \cite{foiq} and \cite{fuSc},
compare also the tables \ref C and \ref{gid} below.
Let us note that isomorphisms between infinite series of coset \cfts\ have
been observed previously. For instance, the $c<1$ minimal conformal
models can be described \cite{goko2} as ${\cal C}[(A_1)_{m-2}^{} \oplus
(A_1)_1^{} \,/\, (A_1)_{m-1}^{}]$, but also as ${\cal C}[
(C_{m-1})_1^{} \,/\, (C_{m-2})_1^{} \oplus (C_1)_1^{}]$; in this case
the field contents is tightly constrained by the representation theory
of the chiral algebra, so that it is relatively easy to construct an
isomorphism as a mapping between primary fields. Our result \erf1
demonstrates for the first time the presence of such isomorphisms for \nn
superconformal theories of arbitrarily high central charge.

The identifications \erf1 are constructed as one-to-one maps between the
primary fields of the respective theories. Both at the level of the
representation of the modular group, and hence for the fractional part
of the conformal dimensions and for the fusion rules, and (by
identifying \rgs s) at the level of the ring of chiral primary fields
we verify that these maps possess the properties needed for an
isomorphism of \cfts.
Clearly one would like to extend the proof from the \furu s
to the full operator product algebra. Because of the technical
difficulties arising in the conformal bootstrap (compare e.g.\
\cite{jf21}), this would be a quite formidable task.  However, it
is reasonable to expect that any two \nn\ super\cfts\ that possess
the same value of the conformal central charge, the same fusion rules,
and the same conformal dimensions modulo integers are in fact isomorphic.
\futnote{In the non-supersymmetric case, examples are known \cite{sche5}
where \cfts\ for which these data coincide are nevertheless distinct
theories. These theories have conformal central charge a multiple of 8
and contain only a single primary field.}
 We are therefore convinced that the two \cts\ in question furnish
merely two different descriptions of one and the same \cft. In this context
note that in general the conformal dimensions of primary fields
change with the `moduli' of some class of \cfts.
For compatibility with the \furu s, the number of primary fields must
then depend on the moduli as well (in fact, when deforming a rational
\cft\ by a massless modulus one generically obtains an irrational theory,
compare the situation at $c=1$).
The arguments in favor of the interpretation of the relations \erf1
as isomorphisms seem to us already conclusive for any
fixed choice of a pair of theories from the list \erf1; they
become even more convincing when one realizes that
our identifications always come in infinite series.

Similar remarks apply to the structure of the chiral ring.
We can substantiate our expectation that there is not only a one-to-one
map between the chiral primary fields of the theories, but that the sets
of chiral primaries also
possess isomorphic ring structures, by various arguments.
First note that the identification of the sets of \rgs s of two \nn
theories implies that they possess the same \pop.
{}From the experience with coset constructions,
the observation that there exist
\cts\ with coinciding \pop s is not very spectacular.
However, it has in fact been shown \cite{sche3,fuSc} that not only
the \opop s,
but also the {\em extended\/} \pop s (introduced in \cite{sche3})
of the relevant theories appearing in \erf1 coincide;
 \futnote{Surprisingly, it seems that in fact for {\em all\/} \nn \cts\
for which the \opop s are identical, the same holds for the \epop s
as well.}
 note that the extended \pop\ describes explicitly part of the structure of the
chiral ring, whereas the ordinary \pop\ essentially counts multiplicities.
Second, the mapping between \rgs s, and thus also between chiral
primary fields, leaves the superconformal charge $q$ invariant.
When proving this, it is important that (in contrast to the case
of generic primary fields of a \ct) for \rgs s not only can we easily
compute the conformal weight exactly (and not just modulo
integers), but also the superconformal charge $q$ \cite{fuSc}.
In addition, the ring product of the chiral ring is highly constrained
by the fusion rules. Namely, since the ring product is defined as the operator
product at coinciding points, the fusion rules (together with naturality
\cite{mose2}) determine which of the structure constants of the chiral
ring are non-zero.
Finally, the charge conjugation on the fusion ring is implemented by the
fusion coefficients ${\cal N}_{i j}^{\ \;0}$, and thus our map respects
charge conjugation, too. In particular, the
charge conjugation behavior of the \rgs s is respected. Since conjugation
on the chiral ring is induced by the ordinary charge conjugation on the
\rgs s via spectral flow (which means that there is a highly non-trivial
interplay between the chiral ring and the representation of the modular
group on the characters), it follows that the map is compatible with the
conjugation of the chiral ring.

As it turns out, the identifications \erf1 are also
interesting in the context of the field identification problem that
arises in coset \cfts. Namely, field identification fixed points
are mapped on non-fixed points, so that the duality provides additional insight
into the procedure of fixed point resolution.
(The resolution procedure for field identification fixed points
shows up in two different ways: for models of $BB$ type, or of $B$ type
with rank
and level odd, fixed points are mapped on longer orbits, while for $B$
type theories
at odd level and even rank the resolution is accomplished by mapping
on pairs of so-called spinor-conjugate orbits.)

The plan of our paper is as follows. The various level-rank dualities
\erf1 of \cts\ are consecutively dealt with in the sections 6 to 9
(the isomorphism statements are made in the equations \erf{BB}, \erf{DD},
\erf{BC}, and \erf {CC}, respectively). These sections make heavy use
of underlying level-rank dualities for the \wzwts\
 \cite{nars,mnrs} the coset models are
composed of.
For the benefit of the reader
we describe the relevant aspects of these dualities in
some detail in the sections 3 to 5, in a formulation that is
adapted to the needs in \nn theories,
making in particular frequent use of modern simple current terminology.
In addition, we present in section 2 a brief reminder about some results
and formul\ae\ from \cft\ that are needed in the sequel.

To conclude this introduction to the subject, let us mention that
level-rank dualities for \nn \cts\ have first been conjectured, for
\hsc s, in \cite{kasu2}; this conjecture just relied on the
symmetry of the conformal central charges of the relevant coset theories.
Calculations of the spectra of \nn \cts\ were first
performed in \cite{foiq,sche3} for \hsc s, and in \cite{fuSc} for non-\hsc s.
The results of \cite{sche3} provided some evidence that the dualities
indeed exist; in particular, it was realized that for $B$ type theories
at odd level and even rank the $D$ type modular invariant must
be used rather than the diagonal one.
In the present paper, we combine the level-rank dualities of \wzwts\
with the properties of
simple current symmetries to construct a map between the primary
fields of the \nn \cts\ in question that makes the level-rank
duality explicit and is expected to be an isomorphism of the two \cfts.
It is worth to stress that the underlying level-rank dualities of
\wzwts\ are definitely not isomorphisms of \cfts.
In particular, these WZW dualities are typically {\em not\/}
mappings between primary fields, but rather between simple current
orbits of (part of) the primary fields. As we will see, this fits
perfectly to the application to \cts, because owing to the necessary
field identifications the physical fields of a \ct\ can be characterized
in terms of combinations of simple current orbits only. In some cases
this technical complication makes
the formulation of the mapping somewhat awkward (and adds to
the length of our paper), but, nonetheless,
the mappings are based on simple current symmetries, and
hence on natural objects of the underlying \wzwts.
We shall show in the sequel that these mappings have the properties
required for isomorphisms of \cfts.

In \cite{kasu2} it was
conjectured that a relation between $B$ type theories at even rank and
even level should exist, too. In this case  non-diagonal modular invariants
must be chosen, but up to now it is not yet clear which of them could
do the job.
 \futnote{Also, none of these \nn models is relevant to
string compactification. For us this is another reason to refrain from
investigating these dualities here.}
 Finally, based on a free field realization of the symmetry algebra,
a level-rank duality for the $A$ type \hsc s has been shown to be present
at the level of symmetry algebras \cite{kasu2}. It would be interesting
to explore these dualities by the techniques developed in the present paper.

\section{Results from \cft} \subsection{Primary fields}

The collection of fields of a \twodim\ \cft\ carries the structure
of a direct sum of irreducible highest weight modules $[\phi_i]$
of the symmetry algebra. The fields corresponding to the highest weights
are the primary fields $\phi_i$.
Upon forming radially ordered products, the fields realize a closed
associative \opa. A large amount of information about the \opa\ is
contained in the \furu s of primary fields, which can be written as formal
products, $\phi_i\star\phi_j=\sum_k\nijk\phi_k$;\, \nijk\ counts the
number of times that $[\phi_k]$ appears in the operator product of
$\phi_i$ and $\phi_j$.

The characters $\chi^{}_i(\tau)={\rm tr}_{[\phi_i]}\exp(2\pi\ii\tau(L_0
-c/24))$ associated to the modules $[\phi_i]$
span a unitary module of the group SL$(2,\zet)$, the
twofold cover of the modular group PSL$(2,\zet)$. This group is
generated freely by elements $S$ and $T$, modulo the relations
$S^2=C=(ST)^3$, where $C^2=\one$; on the characters, these generators act as
$S:\ \tau\mapsto-1/\tau$,\, $T:\ \tau\mapsto\tau+1$. Thus in particular
$T$ is represented as a matrix whose entries are determined by the
fractional part of the conformal dimensions $\Delta_i$ of the
primary fields, $T_{ij}=\delta_{ij}\,
\exp(2\pi\ii(\Delta_i-c/24))$. Further, the \furu s of the theory
can be calculated from the matrix $S$ via the Verlinde formula
\cite{verl2}. The largest eigenvalue of the matrix ${\cal N}_i$ with
elements $({\cal N}_i)_{jk} =\nijk$, the {\em quantum dimension\/} of
$\phi_i$, is of paricular interest; it equals $S_{i0}/S_{00}$, where
the index `0' refers to the identity field \bfe.

Among the primary fields of a \cft\ there may be fields $J$ with the
property that $J\star J^+_{}=\bfe$. These fields are known as
simple currents; the collection of simple currents of a \cft\ forms
an abelian group, with the product given by the fusion product and
inversion given by conjugation. The fusion product of a simple
current with an arbitrary primary field $\phi$ of the \cft\ consists
of only a single primary field, and correspondingly we will often
use the notation $J\phi=J\star\phi$ for this field; $\phi$ and $J\phi$
are said to lie on the same {\em orbit\/} \wrt the simple current $J$.
For simplicity we will also write multiple fusion products of simple
currents as powers,
$J^m=\underbrace {J\star J\star \ldots \star J}_{m\;{\rm factors}}$\,.
Simple currents play an important
role for the construction of non-diagonal modular invariants
and (as we will discuss in subsection 2.3) for the description
of the spectrum of coset \cfts\ \cite{scya5}. In particular, the
so-called $D$ type modular invariants
correspond, roughly speaking, to incorporating a simple current into the
chiral algebra. The $D$ type invariant relevant to us is induced
by a simple current of order two (i.e., $J^2=\bfe)$ with integral
conformal weight. Thus it is of the form $\sum(N_0/N_\phi)\, | \sum_{i=0}
^{N_\phi-1} \chi^{}_{J^i \phi} |^2$,
where $N_\phi$ denotes the length of the simple current orbit containing
the primary field $\phi$ and $N_0=2$ the length of the orbit of the
identity field \bfe, and were the first sum is restricted to orbits that
have vanishing monodromy charge \wrt $J$.

In \nn superconformal field theories another
interesting set of fields are the chiral primary fields. For unitary
field theories (these are the only ones we are going to deal with)
they can be characterized as those fields in the \NS sector for
which the relation $\Delta=q/2$
between the conformal dimension and the superconformal \uE-charge holds.
Their operator product at coinciding points
provides a ring structure different from the one defined by
the \furu s, the so-called chiral ring. The chiral ring plays a
crucial role for many applications of \nn\ theories; for instance,
it encodes interesting phenomenological information when one uses the
theories as the inner sector of a superstring compactification, and also
the relation to topological field theory is mainly through this ring.
Via spectral flow \cite{levw} the chiral ring is in one-to-one correspondence
to the set of ground states of the Ramond sector of the \nn theory,
which in our context is easier to deal with.

\subsection{\wzwts}

A \WZW\ (WZW) theory is a \cft\ whose chiral
symmetry algebra is the semidirect sum of the Virasoro algebra
with an untwisted affine Lie algebra, with the energy-momentum
tensor being quadratic in the currents, i.e.\ in the
generators of the affine algebra.
Many quantities of interest of a \wzwt\ can be described entirely
in terms of the horizontal subalgebra, i.e.\ the simple Lie algebra
$g$ that is generated by the zero mode
currents, and of the level $K$ which (for unitary theories) is
a non-negative integer. For instance, the conformal central charge
is $c(g,K)=K\,{\rm dim}(g)/(K+\gv)$, with \gv\ the dual Coxeter number
of $g$.

The primary fields of a left-right symmetric unitary \wzwt\ are in one-to-one
correspondence with the integrable highest weights, i.e.\ with the
dominant integral weights \L\ of $g$ that satisfy
  \be (\L,\theta)\leq K ,   \label{int} \ee
where $\theta$ is the highest root of $g$ (normalized such that
$(\theta,\theta)=2$). The
conformal dimension of a primary field with highest weight \L\ is
  \be  \hl\equiv\Delta^{}_{(g)}(\L)=\frac{(\L,\L+2\rho)}{2(K+\gv)}, \labl{cdim}
where $\rho$ is the Weyl vector $\rho=\sumin\li$, with \li\ the
fundamental weights of $g$.

The situation is particularly simple for
$g=D_d$ at level one. Then there are four primary fields
corresponding to the singlet (0), vector (v), spinor (s), and
conjugate spinor (c) representation of $D_d$, or, in other words,
to the conjugacy classes of the $D_d$ weight lattice; their conformal
dimension is
  \be  \Delta= \left\{ \begin{array}{lll} 0 & {\rm for} & 0 \,, \\[1 mm]
  1/2 & {\rm for} & {\rm v} \,, \\[1 mm]
  d/8 & {\rm for} & {\rm s,\,c} \,.  \end{array}\right.\labl{cdd}
The modular matrix $S$ of $D_d$ reads
  \be S((D_d)_1) = \onehalf \left( \begin{array}{r r r r}
          1  &  1  &  1  &  1         \\[.3 mm]
          1  &  1  &  -1 &  -1       \\[.3 mm]
          1  &  -1 &  \ii^{-d} & - \ii^{-d} \\[.3 mm]
          1  &  -1 &  -\ii^{-d} & \ii^{-d}  \end{array} \right ) .
  \labl{sD}

Another important class of \cfts\ are those describing a single
free boson compactified on a circle of rational radius squared, to which
for simplicity we will refer as a \wzwt\ with horizontal subalgebra
\uE. The primary fields $\phi_Q$ of these theories are labelled by
\uE-charges\,
$Q\in\{0,1,,...\,,{\cal N}-1\}$, where the number $\cal N$ of
primaries is related to the radius of the circle. (The values
of the integer $\cal N$ that appear in the \nn \cts\ of our interest
will be given in table \ref{nn}.)
The conformal dimension of a \uE-primary of charge $Q$ is $Q^2/2{\cal N}$.
The $S$-matrix elements of a \uE \wzwt\ are
  \be  S_{PQ} = \frac1{\sqrt{{\cal N}}}\, \exp(-2\pi\ii\,PQ/{\cal N}).
  \labl{su}

To fix the notation, let us also list the
simple currents of the \wzwts\ of our interest. For $B$ and $C$
type theories, there is a single simple current besides the identity
primary field; this current will be denoted by $J$ (the corresponding
highest weight is $K\Lambda_{(1)}$ for $B_r$, and
$K\Lambda_{(r)}$ for $C_r$ theories). For $D_r$ type theories, there
are three non-trivial simple currents, corresponding to the
highest weights $K\Lambda_{(1)}$, $K\Lambda_{(r)}$, and $K\Lambda_{(r-1)}$;
they are denoted by \jv, \js, and \jc, as their fusion rules
are isomorphic to the multiplication of
the vector (v), spinor (s), and conjugate spinor (c) conjugacy classes.
Finally, for \uE\ \wzwts, the fusion rules read $\phi_P\star\phi_Q=
\phi_{P+Q\,{\rm mod}\,{\cal N}}$, and hence
any primary field is a simple current.

\subsection{Coset theories}

The idea of the coset construction \cite{goko2} of \cfts\ is to associate
to any pair $g$, $h$ of reductive Lie algebras for which $h$ is a
subalgebra of $g$, a \cft\ called the coset theory
  \be  \cosetK gh \,. \ee
By definition \cite{goko2}, the Virasoro generators of the \ct\
are obtained by subtracting the Virasoro generators of the \wzwt\
based on $h$ from the ones of the \wzwt\ based on $g$. If $g$ is
simple, then the level $K_i$ of any simple summand $h_i$
of $h$ is related to the level $K$ of $g$ by $K_i=I_i K $,
where $I_i$ is the Dynkin index of the embedding $\hing$.

In order to check whether this definition of the coset Virasoro
algebra leads to a well-defined \cft, one also has to specify the
spectrum of primary fields of the theory. As it turns out,
to obtain these primary fields of the \ct\ is a somewhat delicate
issue. To some extent, the field contents can be read off the
so-called branching functions $b^\Lambda_\lambda$, which
are the coefficient functions in the decomposition
  \be \Chi_\Lambda^{}(\tau) = \sum_\lambda b^\Lambda_\lambda(\tau)\,
  \chi_\lambda^{}(\tau)  \labl{:}
of the characters $\Chi_\Lambda$ of $g$ with respect to the characters
$\chi^{}_\lambda$ of $h$.
(Here $\Lambda$ and $\lambda$ stand for integrable highest weights of
$g$ and $h$, respectively, if $g$ and $h$ are simple, and similarly
in the general case.)
The behavior of the branching functions  under modular
transformations suggests that the coset theory associated to the embedding
$h \hookrightarrow g$ might be essentially something like $g\oplus h^*$,
where the notation `*' indicates that the complex conjugates of the
modular transformation matrices of the \wzwt\ based on $h$ should
be used. Note that if $S$ and $T$ generate a representation of the modular
group, the same is true for $S^*$ and $T^*$. If there exists a \cft\
whose characters transform according to this complex conjugate
representation, it is called the {\em complement\/} of the $h$
theory \cite{scya6}.

However, closer inspection shows that the coset theory is in fact
rather different from $g\oplus h^*$.
Namely, in a \ct\ the requirement that the
characters span a unitary module of SL$(2,\zet)$
forces us to associate physical fields not with individual
branching functions, but rather with certain equivalence classes of them.
This is commonly referred to as `field identification'
\cite{gepn8,scya5}.
As already mentioned, the field identification can be understood in
terms of simple current symmetries. Namely \cite{scya5}, technically it is
convenient to implement the identification procedure by means of
the action of appropriate simple currents, known as
identification currents. These are specific tensor products, to be
denoted as $(J_{(g)}^{}\,/\,J_{(h)}^{})$, of the simple currents of the
\wzwts\ that underly the \ct.

\begin{table}[t]
\caption{Some \nn superconformal \cts\ and their Virasoro charges}
\label{C} \begin{center}
{\small \begin{tabular}{|l|l|r|} \hline&&\\[-3 mm]
\multicolumn{1}{|c|}{name} & \multicolumn{1}{c|}{\coset{g_K^{}\op(D_d)
_1^{}}{\bigoplus_i(h_i)^{}_{K_i}\oplus(\uE)^{}_{\cal N}}{} } &
\multicolumn{1}{c|}{$c$} \dhlinz{3.7}  \label{nn}
($B,2n+1,K$) & \CSfull{B_{n+1}}{2n+1}{B_n}{K+2}{4(K+2n+1)}
  & $\displaystyle\frac{3K(2n+1)}{K+2n+1}$  \\ &&\\[-3 mm]
($B,2n,K$)   & \CSfull{D_{n+1}}{2n}{D_n}{K+2}{4(K+2n)}
  & $\displaystyle\frac{6Kn}{K+2n}$   \\ &&\\[-3 mm]
($BB, 3, K$) & \CSfulld{B_3}{7}{A_1}{2K+8}{A_1}{K+3}{2(K+5)}
  & $21-\displaystyle\frac{96}{K+5}$   \\ &&\\[-3 mm]
($BB, n, K$)\,,$\;${\footnotesize $n\!>\!3\!\!$} &
    ${\cal C}[(B_n)_K\op(D_{4n-5})_1\,/\,$ & \\[-1 mm] &$ \qquad\qquad
              (B_{n-2})_{K+4}\op(A_1)_{K+2n-3}\op({\rm u}_1)_{2(K+2n-1)}]$
  & $12n-15-\displaystyle\frac{24(n\!-\!1)^2}{K\!+\!2n\!-\!1}$ \\ &&\\[-3 mm]
($CC, n, K$) & \CSfull{C_n}{2n-1}{C_{n-1}}{K+1}{2(K+n+1)}
  & $6n-3-\displaystyle\frac{6n^2}{K+n+1}$
       \\[2.3 mm] \hline \end{tabular} }
  \end{center} \end{table}

As long as all orbits with respect to the identification currents
have equal size, the orbits are precisely the independent physical
fields we are after.The situation is more involved if the orbits have
different numbers of representatives. These numbers are
divisors of the length $N_0$ of the orbit of the identity field.
Orbits with less than $N_0$ representatives are referred to
as `fixed points' of the identification currents.
Among the theories of our interest, only
the cosets of $CC$ type and $(B,2n,2k+1)$ do not possess any fixed points.
If fixed points are present, one has to complement the previous
prescription for finding the physical fields by a so-called
`fixed point resolution.' Every fixed point of length $N_f$ has to
be resolved in $N_0 / N_f$ distinct physical fields.

As has been shown in \cite{kasu}, \cts\ \coset{\tilde g}{\tilde h}K with
  \be  \tilde g=g\oplus D_d \qquad \tilde h=h\oplus\uE  \ee
with $2d={\rm dim}\,g-{\rm dim}\,\tilde h$ and $D_d$ at level one
can possess \nn superconformal symmetry; further, all combinations of
$g$ and $h$ for which this happens have been listed \cite{kasu,schW}.
In table \ref C we present those cases which are relevant for our
present purposes.

\begin{table}[t] \caption{Identification currents for \nn \cts} \label{gid}
\begin{center} \begin{tabular}{|l|c|l|} \hline &&\\[-3.2 mm]
\multicolumn{1}{c|}{name} & \multicolumn{1}{|c|}{$N_0$}
& \multicolumn{1}{|c|}{Independent identification currents} \dhlinz{3.7}
($B,2n+1,2k+1$) & 4 & \arrayleft{ J_{(1)}:=(J, 0 \csp J, 0) \\[.3 mm]
                             J_{(2)}:=(J,\jv\csp 0, \pm4(k+n+1))}
               \\[-3.2 mm] &&\\
($B,2n+1,2k)_{\mid D}$ & 8 & \arrayleft{ J_{(1)}:=(J, 0 \csp J, 0) \\[.3 mm]
                             J_{(2)}:=(J,\jv\csp 0, \pm2(2k+2n+1))\\[.3 mm]
                             J_{(3)}:=(J,0\csp 0, 0))}
               \\[-3.2 mm] &&\\
($B,2n,K$)   & 8 & \arrayleft{ J_{(1)}:=(\jv,0\csp\jv,0) \\[.3 mm]
                             J_{(2)}:=(\js,(\jv)^n \csp \js,(K+2n))}
               \\[-3.2 mm] &&\\
($BB,n,K$) & 4 & \arrayleft{ J_{(1)}:=(J, 0 \csp J, 0, 0) \\[.3 mm]
                             J_{(2)}:=(J, 0 \csp 0, J, \pm (K+2n-1))}
               \\[-3.2 mm] &&\\
($CC,n,K$) & 2 & $\quad\, J_{(1)}:=(J, (\jv)^n \csp J, \pm(K+n+1))$
\\[1.6 mm] \hline \end{tabular} \end{center} \end{table}

For \nn \cts, the fixed point resolution procedure
has been worked out in \cite{sche3} (for the so-called \hsc s) and
in \cite{fuSc} (for non-hermitian symmetric \nn theories).
The primary fields $\Phi$ of a \nn \ct\ \coset{g\oplus D_d}{h\oplus\uE}K
may be labelled by the weights carried by
the primaries of the \wzwts\ it is composed of, i.e.\
  \be  \Phi \heq (\Lambda,{\rm x}\csp\lambda,Q)  \labl P
with $\Lambda$ and $\lambda$ integrable highest weights of the $g$
and $h$ algebras, x a conjugacy class of $D_d$,
and $Q\in\{0,1,,...\,,{\cal N}-1\}$ a \uE-charge. However, as a
consequence of the necessary field identification, this
labelling is not one-to-one. Rather, all combinations of labels
that are connected via the action of the identification currents
describe one and the same primary field; moreover, fixed points
have to be resolved, which introduces an additional label $i$
according to
  \be  \Phi_{\rm fix} \heq (\Lambda,{\rm x}\csp\lambda,Q)_i^{}  \labl i
The identification currents (including the simple current that
implements the $D$ type modular invariant in the case of
($B,2n+1,2k)$) of the \nn theories of table \ref C
are displayed in table \ref{gid}.
 \futnote{Non-trivial simple currents of \wzwts\ are denoted by
the symbols $J$, \jv\ etc.\ introduced above.\\
We also note that in \cite{sche3}
a different value for $\cal N$ is used
for $(B,2n,K)$, namely ${\cal N}=16(K+2n)$. This is compensated by
adding, as is done in \cite{sche3}, a further identification current
which is trivial in
all parts except the \uE\ part, namely $(1,1\csp 1, \pm 8(K+2n))$.}

The conformal dimension of the field $\Phi$ is modulo integers
 \be  \Delta(\Phi) = \Delta_{(g)}^{}(\L)+\Delta^{}_{(D_d)}({\rm x})
 -\Delta_{(h)}^{}(\lambda)-\Delta^{}_{({\rm u}_1)}(Q) , \labl h
where $\Delta_{(g)}^{}(\L)$ and $\Delta_{(h)}^{}(\lambda)$ are
defined as in \erf{cdim}, $\Delta_{(D_d)}({\rm x})$ is given in
\erf{cdd}, and $\Delta_{(1)}(Q)=Q^2/2{\cal N}$.
The superconformal charge $q$ is modulo 2 given by
  \be  q(\Phi) = \sum_\alpha {\rm x}^\alpha - \frac{\xi Q}{K+\gv} \,. \labl q
Here ${\rm x}^\alpha$ are the components of x in the orthonormal basis of
the $D_d$ weight space, and $\xi$ is a rational number
that can be calculated \cite{fuSc} with
the help of the relation between the normalization of the \uE-charge
and the length of $\rhog-\rhoh$. For the theories of our interest, one
has $\xi=n$ for $(B,2n,K)$ and $(CC,n,K)$,
$\,\xi=n+\onehalf$ for $(B,2n+1,K)$, and $\,\xi=2(n-1)$ for $(BB,n,K)$.

For the \smat\ elements in a coset theory with fixed points one makes the
ansatz \cite{sche3}
  \be \tilde{S}_{e_i f_j}
  = \frac{N_e N_f}{N_0} S_{ef} + \Gamma^{ef}_{ij}, \labl{fc}
where $i = 1,2, \ldots,N_0/N_e$ and $j = 1,2, \ldots,N_0/N_f$
label the fields into which the naive fields $\phi_e$ and $\phi_f$ are
to be resolved if they are fixed points.
Modular invariance implies the sum rule
  \be  \sum_i \Gamma^{ef}_{ij}  = 0 = \sum_j \Gamma^{ef}_{i j} \labl S
for the $S$-matrix elements of the fields $\phi_{f_i}$.
Note that if either $\phi_e$ or $\phi_f$ is not a fixed point, the sum
rule tells us that $\Gamma$ vanishes. So $\Gamma$ is non-zero only for
pairs of resolved fixed points, in which case it has also to be symmetric
under simultaneous
exchange of $(e, i)$ and $(f, j)$, since the total \smat\ $\tilde S$
must have this property.

To find a solution for $\Gamma$ for \nn \cts,
we assume that with respect to the individual
entries of the multi-index $(f,i) = (\Lambda,{\rm x}\csp\lambda,Q)_i$\,,
$\,\Gamma$ factorizes as
  \be  \Gamma_{i;j}^{\Lambda,{\rm x},\lambda,Q; \Lambda',{\rm x}'
  ,\lambda',Q'}  =  \Gamma_{(g)}^{\Lambda\,\Lambda'} \Gamma_{(D_d)}^
  {{\rm x\,x}'} \Gamma_{(h)}^{\lambda\,\lambda'} \Gamma_{({\rm u}_1)}^
  {Q\,Q'} P_{ij} \,, \labl{fac}
where
  \be P_{ij} = \delta_{ij} - \frac{N_f}{N_0}    \ee
if $N_e=N_f$. (For brevity we will focus here on fixed points \wrt
identification
currents of prime order, in which case $N_e=N_f=1$ for all fixed points.
If the order is not prime, an iterative procedure \cite{scya6} must be
applied to resolve fixed points with $N_f>1$. Also note that for
$N_0=2$ and $N_f=1$ the factorization of $P$ in \erf{fac} is a necessary
consequence of the sum rule \erf S.) In \cite{scya6} a
general prescription to find the matrix factors $\Gamma_{(\cdot)}$
for \wzwts\ was given. These matrices, as well as
the character modifications to be described below,
can be viewed as the $S$-matrix and characters of some \cft,
which is called the `fixed point theory.'
In most cases the fixed point theory turns out to be another
\wzwt. As we shall see
below, this prescription to deal with fixed points is consistent with
our level-rank duality in \cts. In fact, level-rank duality will even provide
additional insight in the nature of the fixed point resolution.

Resolving a fixed point amounts to considering fields having different
characters $\chi^{}_{f_i}$, i.e.\ the naive branching function
$\chi_f$ of the `unresolved fixed point' must be modified by adding
an appropriate multiple of a character \ckri\ of the fixed point theory.
Again, modular invariance implies a sum rule, namely
  \be  \sum_i \chi^{}_{f_i}  = \frac{N_0}{N_f}\, \chi^{}_f \,. \labl"
For $N_0/N_f=2$, we denote the two resolved
fixed points by $f_{\pm}$. According to \erf", the corresponding
characters read
  \be \chi_{f_{\pm}}^{} = \chi_f^{} \pm v \ckri \,,    \labl{pm}
where without loss of generality we can assume that $v>0$.
It is easy to see that these characters transform indeed according to the
resolved \smat\ \erf{fc}, with $P_{++}=P_{--} = \onehalf$ and $P_{+-}
=P_{-+} = - \onehalf $.

In order to identify the chiral rings of \nn \cts, we will look
at the \rgs s. Any \rgs\ has at least one representative
  \be \Phi_R^{}=(\Lambda,{\rm x},\tilde\lambda)    \labl{Ltw}
for which \L\ and \latw\
are related through a Weyl group element $w\in W(g)$ according to \cite{levw}
  \be \latw + \rhoh = w(\Lambda + \rhog ) \,.  \labl{ltw}
Here \latw\ incorporates both the weight $\lambda$ of the semi-simple part
$h$ of $\tilde h$ and the \uE-charge $Q$, and
  \be  {\rm x}=\left\{ \begin{array}{ll} {\rm s}& \mfor\ \sign(w)=1
  \,,\\[2.1mm] {\rm c}& \mfor\ \sign(w)=-1 \,. \end{array}
  \right.\labl c
Furthermore, the Weyl group element $w$ has to be chosen in such a manner
that $\lambda$ is a highest weight of $h$
(this fixes uniquely one representative of each element of
the coset $W(g)/W(h)$).

The superconformal charge $q$ (including the integer part) of a \rgs\
is conveniently computed from the formula \cite{fuSc}
  \be  q(\Phi_R) = \frac d2 - l(w) - \frac{\xi Q}{K + \gv}  \labl{Len}
that relates $q$ to the \uE-charge $Q$ and to the length $l(w)$ of the
Weyl group element $w$ that appears in \erf{ltw}
($\xi$ is the number introduced in \erf q).
The length $l(w)$ can be obtained \cite{HUmp2} as the number of
negative roots $\alpha$ of $g$ for which $w(\alpha)$ is a positive root,
  \be  l(w)=\;\mid\!\{\,\alpha\!<\!0\mid w(\alpha)\!>\!0\,\}\!\mid \,.
  \labl{HUmp}

For an \nn \ct\ \coset{g\oplus D_d}{h\oplus\uE}K without fixed
points, the number of chiral primary fields is \cite{levw}
  \be \mu = \frac{N(g)}{|Z(g)|}\, \frac{|W(g)|}{|W(h)|}  \,, \labl{levw}
where $N$ is the number of primary fields of the \wzwt\ based on
$g$ at level $K$, and $Z(g)$ is
the center of the universal covering group whose Lie algebra
is $g$ (which is isomorphic to the group of simple currents of
the \wzwt). The factor $1/|Z(g)|$ takes care of the necessary field
identifications among representatives of the form \erf{Ltw}, \erf{ltw}.
In contrast, if an \nn \ct\ has fixed points, the number
of \rgs s is larger than \erf{levw}. Namely, each primary field of
$g$ still gives rise to $|W(g)/|W(h)|$ representatives of
chiral primaries, but in addition for fixed points it is still true
that (after resolution of fixed points) every
\rgs\ has a representative whose $g$- and $h$-weights fulfill \erf{ltw},
and that every equivalence class containing one
representative with $\tilde \lambda = w( \Lambda + \rhog ) - \rhoh$
yields precisely one \rgs .

\sect{$B$ type \wzwts\ at odd level} \label{secbodd}

In this section we will describe a map $\tau$
between the \wzwts\ \bnk\ and \bkn\
that has simple behavior \wrt the modular matrices $T$ (i.e.,
\wrt \cdim s modulo integers) and $S$.
Thus the two theories that are connected by $\tau$ are related by
exchanging twice the rank plus one (recall that $B_n\cong {\rm so}(2n+1)$)
with the level of a $B$ type affine Lie \alg;
a relation of this type is called {\em level-rank duality}.
As mentioned in the introduction, such dualities emerge in
various different contexts; here we will concentrate on those
aspects that are needed for the identifications of \nn \cts\ in
sections \ref{sBB} to 8 below. The level-rank duality in question
was first realized in \cite{mnrs};
in the notation of \cite{mnrs},
our map $\tau$ corresponds to the map `tilde' for $B$ weights that
are tensors, and to the map `hat' for spinor weights, respectively.
 \futnote{We are grateful to the authors of \cite{mnrs} for extensive
explanation of their notation.}
To be more precise, $\tau$ will be a
one-to-one map between orbits \wrt the relevant simple currents $J$
of the two theories. Thus, to start, we note that the number of primaries
of the \bnk\ \wzwt, i.e.\ the number of integrable representations
of the affinization of $B_n$ at level $2k+1$, is
  \be  N^B_{n,2k+1}=\sum_{l=0}^k \biN{2k-2l+3}{2} \biN{l+n-3}{l}
  =(4k+3n)(n+k-1)!/n!k! \,; \labl{nB}
of these,
  \be  F^B_{n,2k+1}=\bin{n+k-1}{k} \labl{fB}
are fixed points, so that the number of orbits is
$2\biN{n+k}{k}$. This is invariant under $n\leftrightarrow k$, so
that indeed a one-to-one map between the respective sets of
orbits is conceivable.

For any integrable highest weight $\L=\sumin \L^i\li$ of \bnk,
denote by
  \be  \cl=\L^n \bmod 2 \ee
the conjugacy class of \L. For brevity, we will often refer to
\L\ as a `tensor' and as
`spinor' weight if $\cl=0$ and $\cl=1$, respectively.
Consider now the components of \L\
in the orthonormal basis of the weight space; they read
  \be  \eli(\L)=\sum_{j=i}^{n-1}\L^j+\onehalf\,\L^n.  \labl{22}
Adding to these numbers the components of the Weyl vector
as well as a term $\onehalf\,(1-\cl)$ such as to make the result
integer-valued, one defines
  \be  \tli(\L):=\eli(\Lambda+\rho)+\onehalf\,(1-\cl)
  =\sum_{j=i}^{n-1}\L^j+n+1-i+\onehalf\,(\L^n-\cl).  \labl{23}
Under the action of the simple current $J$ that carries the
highest weight $(2k+1)\Lambda_{(1)}$, the numbers
\eli, $i=2,3,...\,,n$, are invariant, while \ele\ gets replaced
by $2k+1-\ele$. As a consequence, we may characterize any orbit
of $J$  by a set of $n$ positive integers \tli, $i=1,2,...\,,n$
subject to $\tli>\tilde\ell_j$ for $i<j$ as well as $\tle\leq k+n$,
or in other words, by a subset \ml\ of cardinality $|\ml|=n$ of the set
  \be  M:=\{1,2,\ldots,k+n\} \,.  \labl m
Each such subset describes precisely one tensor and one spinor orbit
(in particular, there are as many spinor orbits as tensor orbits if the
level of $B_n$ is odd), and conversely, any integrable
highest weight of \bnk\ corresponds to precisely one of these subsets.

We are now in a position to present the map $\tau$. First consider
spinor weights \L\ of \bnk. Given the associated subset $\ml\subset M$,
define the complementary set
  \be  \{\Tli\}\equiv\mtl:=M\setminus\ml, \labl{taus}
where the numbers \Tli\ are to be ordered according to
$\Tli>\tilde\ell_j^{(\tau)}$ for $i<j$.
Since this subset of $M$ again satisfies $\Tle\leq k+n$,
and is of cardinality $k$, it describes precisely one orbit
$\{\tl, J \star\tl\}$ of integrable highest spinor weights of \bkn.
Also note that \ml\ describes a spinor fixed point iff $k+n\in\ml$
(in contrast, there do not exist tensor fixed points at odd level); thus
spinor fixed points are mapped to spinor orbits of size two, and vice versa.

Let us now check how the modular matrix $T$ transforms under the map $\tau$.
By combining the formul\ae\ \erf{cdim} and \erf{taus}, and
inserting the strange formula for the length of the Weyl vectors,
one finds
  \be  \begin{array}{ll}  \hl+\htl &=\; [ \sum_{j\in\mL}j^2-(\rho,\rho)
  +\sum_{j\in\mtL}j^2-(\rhot,\rhot)] / [4(k+n)]  \\ {}\\[-2.5 mm]
  &=\; [ \sum_{j=1}^{k+n}j^2-\frac1{12}\,(4 n^3-n+ 4 k^3-k)] /
  [ 4(k+n)]  \\ {}\\[-2.5 mm]
  &=\; \frac18\,(k+n+2kn+\frac12) \,,  \end{array}\ee
where $\rho$ and \rhot\ denote the Weyl vectors of $B_n$ and $B_k$
respectively. (Recall that we choose the representatives \L\ and
\L$^{(\tau)}$ such that $\tle\leq k+n$ and $\tle^{(\tau)}\leq k+n$;
as the \cdim s of the elements of a spinor orbit differ by an
integer, this means that for the other member of a length-two orbit, the
formula holds true modulo \zet).

For tensors we will have to consider a
definition of $\tau$ that is different from that for spinors
\cite{mnrs}. Namely, while
again the complement of \ml\ in $M$ plays a role, we now define \mtl\ by
  \be  \{\Tli\}\equiv\mtl:=\{k+n+1-l\mid l\in M\setminus\ml\}. \labl{taut}
By definition, this maps tensor orbits to tensor orbits, and again the
image covers all such orbits of \bkn\ precisely once.
For the sum of \cdim s we now obtain
  \be  \begin{array}{ll}  \hl+\htl &=\; [ \sum_{j\in\mL}(j-\onehalf)^2
  -(\rho,\rho) +\sum_{j\in M\setminus\mL}(k+n+\onehalf-j)^2
  \\ {}\\[-2.5 mm] & \hsp{51} -(\rhot,\rhot)] \,/ [ 4(k+n)]  \\
 \label{310} {}\\[-2.5 mm]
  &=\; \frac14\,k(k+2n+1)-\frac12\sum_{j\in M\setminus\mL}j \,,  \end{array}\ee
which is a half integer.
(Again this result is true for \L\ such that $\tle\leq k+n$, and
analogously for \tl; the \cdim s of the elements of a tensor orbit differ
by \onehalf\ plus an integer, so that for the other members of the
orbits, the formula still holds modulo \zet/2).

One can visualize the map $\tau$ in terms of Young tableaux $Y(\Lambda)$,
defined as having $\eli(\L)-\onehalf c_\Lambda$ boxes in the $i$th row. The
prescription \erf{taus} corresponds to forming the complement \wrt the
rectangular Young tableau $Y(k\la n)$, followed by reflection at an
axis perpendicular to the main diagonal. Similarly,
the map \erf{taut} corresponds just to reflection at the main diagonal.
For example, consider the following mapping between tensor orbits of
the (self-dual) $(B_3)_7$ \wzwt\
(for better readability, we display, with dotted lines, also
the missing boxes that are needed to extend a tableau $Y(\L)$
to $Y(k\la n)$):
  \be  \bP{60}{60}00   \hdx003 \hbX0{20}2 \fdx{40}{20} \hbX0{40}3 \eP \quad
  \bP{60}{60}00   \mbx{30}{30}{$\longleftrightarrow$} \eP \quad
  \bP{60}{60}00   \vdx{40}03 \vbX003 \fdx{20}0 \vbX{20}{20}2 \eP
  \ee
According to the previous prescriptions, the corresponding orbits
are $\{(1,2,0),(2,2,0)\}$ for the left hand side, and
$\{(0,1,2),(3,1,2)\}$ for the right hand side (here we write the
weights in the basis of fundamental highest weights), and indeed these orbits
are mapped onto each other by \erf{taut}. Considering, instead, the
left hand side as a Young tableau for a spinor orbit, namely for
the fixed point $(1,2,1)$, it gets mapped via \erf{taus} to the
spinor orbit $\{(1,0,3),(3,0,3)\}$, i.e.
  \be \bP{60}{60}00   \hdx003 \hbX0{20}2 \fdx{40}{20} \hbX0{40}3 \eP \quad
  \bP{60}{60}00   \mbx{30}{30}{$\longleftrightarrow$} \eP \quad
  \bP{60}{60}00   \vdx{40}03 \vbX003 \vdx{20}02 \fbX{20}{40} \eP \ee
 \clearpag
As further examples, consider the mappings
  \be \bP{80}{80}00 \hdx0{10}4 \hbX0{30}2 \hdx{40}{30}2 \hbX0{50}3
  \fdx{60}{50} \eP \quad \bP{60}{80}00   \mbx{30}{40}{$\longleftrightarrow$}
  \eP \quad \bP{60}{80}00 \vdx{40}04 \vbX0{20}3 \vdx{20}02 \fdx00
  \vbX{20}{40}2 \eP \ee
and
  \be \bP{80}{80}00 \hdx0{10}4 \hbX0{30}2 \hdx{40}{30}2 \hbX0{50}3
  \fdx{60}{50} \eP \quad \bP{60}{80}00 \mbx{30}{40}{$\longleftrightarrow$}
  \eP \quad \bP{60}{80}00 \vdx{40}03 \vbX004 \vdx{20}02 \fbX{40}{60}
  \vbX{20}{40}2 \eP \ee
between orbits of $(B_3)_9$ (left) and $(B_4)_7$ (right). The first
of these corresponds to the tensor orbits
$\{(1,2,0),(4,2,0)\}\leftrightarrow\{(0,1,1,0),(3,1,1,0)\}$, and the second
to the spinor orbits
$\{(1,2,1),(3,2,1)\}\leftrightarrow\{(1,1,0,3)\}$.

Above, we have already obtained all information that we need
about the modular matrix $T$. Next we want to determine the behavior of
the \smat\ under the map $\tau$. We first recall that
the Weyl group $W$ of $B_n$ acts in the orthonormal basis by
all possible permutations and sign changes of the components.
This implies that
  \be  \sumW{\rm sign}\,(w)\,\exp \mbox{\large[} \frac{\pi\ii}{k+n}\,
  (w(\L+\rho),\Lambda'+\rho) \mbox{\large]} = (2\ii)^n\, \detij
  \mijll , \ee
where
  \be  \mijll:=\sin \mbox{\large[} \frac{\pi\,\eli(\L + \rho )
  \,\ell_j(\Lambda' + \rho )}{k+n} \mbox{\large]} \,. \ee
Inserting this identity into the \kpf\ \cite{kape3,kawa} for the \smat,
one arrives at
  \be  \sll=(-1)^{n(n-1)/2} 2^{n/2-1} (k+n)^{-n/2} \,\detij\mijll.
  \labl{sll}
Now of course this result for the \smat\ refers to particular
highest weights \L\ and $\Lambda'$. However,
what we really would like to compare are not the \smat\ elements
for individual weights, but
\smat\ elements for orbits \wrt simple currents. Now within an orbit,
the sign of $S$ depends on the choice of the representative (except if only
tensor weights are involved). Thus if we want to interpret \erf{sll}
as an equation for orbits, we have to keep in mind that when evaluating
the equation we have to employ specific representatives (namely, those
with the smaller value of $\ell_1$). For the application to \cts\ it
will be crucial that the sign in \erf{sll} is correlated with the
alternative whether the relation \erf{310} between conformal weights
holds exactly or only modulo $\onehalf\zet$.

An analogous computation as for \erf{sll} yields
  \be  \stll=(-1)^{k(k-1)/2} 2^{k/2-1} (k+n)^{-k/2} \,\detij\mijtll
  \labl{stll}
with
  \be  \mijtll:=\sin \mbox{\large[} \frac{\pi\,\elI(\L^{(\tau )} + \rhot )
  \,\ell^{(\tau)}_j(\Lambda'^{(\tau )} + \rhot )}{k+n} \mbox{\large]} \,. \ee
 \clearpag
To relate the numbers \erf{sll} and \erf{stll}, we first note that
\mijll\ can be viewed as a $n\times n$ sub-matrix of the $(k+n)\times
(k+n)$ matrix
  \be  \Om_{ij}:= \left\{ \begin{array}{lll}
  \omij tt:=\sin[(\pi\,(i-\onehalf)(j-\onehalf)/(k+n)]
  & {\rm for} & \cl=\clp=0 \, , \\[1.2 mm]
  \omij ts:=\sin[(\pi\,(i-\onehalf)j)/(k+n)]
  & {\rm for} & \cl=0,\ \clp=1 \, , \\[1.2 mm]
  \omij st:=\sin[(\pi\,i(j-\onehalf)/(k+n)]
  & {\rm for} & \cl=1,\ \clp=0 \, , \\[1.2 mm]
  \omij ss:=\sin[(\pi\,ij)/(k+n)]
  & {\rm for} & \cl=\clp=1 \, , \end{array}\right. \labl{Om}
$i,j=1,2,...\,,k+n$. Similarly, \mijtll\ is a $k\times k$ sub-matrix of
  \be  \tilde{\!\Om}\,_{ij}:= \left\{ \begin{array}{lll}
  \sin[(\pi\,(k+n+\onehalf-i)(k+n+\onehalf-j))/(k+n)]
  \\[.9 mm]  \hsp{39} =(-1)^{i+j+k+n+1}\,\omij tt
  & {\rm for} & \cl=\clp=0 \, , \\[1.2 mm]
  \sin[(\pi\,(k+n+\onehalf-i)j)/(k+n)]=(-1)^{j+1}\,\omij ts
  & {\rm for} & \cl=0,\ \clp=1 \, , \\[1.2 mm]
  \sin[(\pi\,i(k+n+\onehalf-j))/(k+n)]=(-1)^{i+1}\,\omij st
  & {\rm for} & \cl=1,\ \clp=0 \, , \\[1.2 mm]
  \omij ss & {\rm for} & \cl=\clp=1 \, . \end{array}\right. \ee

More precisely, the two submatrices are such that together
they cover each value of $i$ and $j$ precisely once.
As a consequence, one can
use (a simple case of) the so-called Jacobi-theorem
\cite{JEje,mnrs} to relate \sll\ to \stll.
The theorem states that for any invertible matrix $\Om$ whose
rows and columns are labelled by a set $H$,
one has for $I,\,J\subset H$ with $I\cup J=H$, $I\cap J=\emptyset$,
that
  \be  {\rm det}\,[(\Om^{-1})^t_{}]^{}_{IJ} =(-1)^{\Sigma_I+
  \Sigma_J}({\rm det}\,\Om)^{-1}\,({\rm det}\,\Om)_{\Ibar\Jbar}
  \ee
with $\Ibar=H\setminus I$, $\Jbar=H\setminus J$, and
  \be  \Sigma_I=\sum_{j\in I}j, \quad \Sigma_J=\sum_{j\in J}j. \labl{IJ}
Writing
  $\sll=\alpha\, {\rm det}\,\Om_{IJ}, \;
  \stll=\beta\, {\rm det}\,\Om_{\Ibar\Jbar}$ % \labl{asS}
and
  ${\rm det}\,[(\Om^{-1})^t_{}]^{}_{IJ} =\delta\, {\rm det}\,
  \Om_{IJ},$ % \labl{ass}
application of this theorem yields
  \be  \sll=(-1)^{\Sigma_I+\Sigma_J} \alpha\,(\beta\gamma\delta)^{-1}
  \stll  \labl{sst}
with $I=\ml$, $J=M_{\Lambda'}$, and $\Om$ as defined in \erf{Om}.
(Actually, the definition of $\delta$ implies the assumption that
${\rm det}\,\Om_{IJ}
\neq0$ for all choices of $I$ and $J$. This turns out to be true
for all cases we are interested in. Moreover, in some cases in fact
$\delta$ does not depend on the choice of $I$ and $J$ at all.)

An explicit expression for
the number $\alpha$ can be read off \erf{sll}, while when
determining the parameters $\beta,\,\gamma,\,\delta$, one has to
distinguish between tensors and spinors. If both \L\ and $\Lambda'$
are tensors, then by straightforward calculation one finds
  \be  \begin{array}{l}  \beta=(-1)^{k(k-1)/2} (-1)^{k(k+n+1)
  +\Sigma_{\Ibar}+\Sigma_{\Jbar}}\, 2^{k/2-1} (k+n)^{-k/2},
  \\[1.2 mm]  \gamma=(-1)^{(k+n)(k+n-1)/2} ((k+n)/2)^{(k+n)/2},
  \qquad      \delta=(2/(k+n))^n \,.   \end{array}\ee
When inserted into \erf{sst}, this yields, upon use of the identity
$\Sigma_{\Ibar}+\Sigma_I=\sum_{j=1}^{k+n}j=(k+n)(k+n+1)/2$ \cite{mnrs},
  \be  \sll=\stll . \ee
Note that this implies that $\tau$ connects tensor orbits with
identical quantum
dimension. (Since simple currents have quantum dimension 1 and
quantum dimensions behave multiplicatively under the fusion product, the
quantum dimension is constant on simple current orbits.)

If \L\ is a tensor and $\Lambda'$ a spinor, one obtains
 \futnote{Notice that if \L\ is a tensor, then the order of the rows
of $\tilde{\!\Om}_{ij}$ is actually to be read backwards such as to
satisfy the requirement
that the numbers obey $\Tli>\tilde\ell_j^{(\tau)}$ for $i<j$;
this contributes a factor
$(-1)^{k(k-1)/2}$ to $\beta$. If $\Lambda'$ is a tensor, the same factor
arises from an analogous re-ordering of columns. In particular,
for both \L\ and $\Lambda'$ tensors, these factors cancel out.}
  \be  \begin{array}{l}  \beta= (-1)^{k
  +\Sigma_{\Ibar}} 2^{k/2-1} (k+n)^{-k/2},
  \\[1.2 mm]  \gamma=(-1)^{(k+n)(k+n-1)/2}\, 2^{(1-k-n)/2} (k+n)^{(k+n)/2}
  , \qquad    \delta=2^{n-\fl}\,(k+n)^{-n} \,,   \end{array}\ee
where
  \be  \fl:=\left\{ \begin{array}{lll} 1 & {\rm for} & \Lambda' \mbox
  {~\,a fixed point,} \\[1.2 mm] 0 & {\rm for} & \Lambda' \mbox{~\,an
  orbit of length two}. \end{array}\right. \labl{deff}
Thus in this case \cite{mnrs}
  \be  \sll=(-1)^{\Sigma_I+n(n+1)/2}\,2^{\fl-1/2}\,\stll. \labl{sts}
(Again, the sign depends on the choice of representative of the
tensor orbit. It is as given in \erf{sts} if the
representative with smaller value of $\ele$ and $\tle$ is taken.)
In particular, for spinors the quantum dimensions of the orbits of $\Lambda$
and $\tau(\Lambda)$
differ by a factor $\sqrt2 $ for orbits of length 2, and by a factor
$1/\sqrt2$ for fixed points.

Analogously, for \L\ a spinor and $\Lambda'$ a tensor, one obtains
\erf{sts} with $\Sigma_J$ replaced by $\Sigma_I$ and $\fl$ replaced
by $f(\L)$. Finally, if both \L\ and $\Lambda'$ are spinors, we
again have to distinguish between several cases.
Observing that \L\ is a fixed point  iff $k+n\in\ml$, and that
$\Om^{(\rm ss)}_{j,k+n}=\Om^{(\rm ss)}_{k+n,j}
=\sin(\pi j)=0$, we conclude that
  \be  \sll=\stll=0 \ee
if \L\ is a fixed point and $\Lambda'$ belongs to a length-two spinor orbit,
or vice versa. In contrast, if both \L\ and $\Lambda'$ are fixed
points, \sll\ vanishes but \stll\ does not, and the other way round
for both \L\ and $\Lambda'$ belonging to length-two spinor orbits.

\sect{$B$ type theories at even level versus $D$ type at odd level}

In this section we present a map $\tau$ relating \bke\ and \dnk\ that behaves
similarly as the one described in the previous section. However,
for \bke\ we now have to restrict ourselves to tensor weights;
for these, we define \eli\ and \tli\ as
in \erf{22} and \erf{23}. In contrast to odd level, now
the map is no longer one-to-one on the simple current orbits. Rather,
some of the orbits of \bke\ (namely, those tensors which are fixed
points; in contrast to odd level, fixed points now must be tensors)
correspond to two distinct orbits of \dnk.

For \dnk, the components of a weight $\Lambda$ in the orthonormal
basis are
  \be \begin{array}{l}  \eli(\L)=\sum_{j=i}^{n-2}\L^j+\onehalf\,
  (\L^{n-1}+\L^n) \mbox{~~~for~}i=1,2,...\,,n-2\,, \\[2.3 mm]
  \ell_{n-1}=\onehalf\,(\L^{n-1}+\L^n)\,, \qquad
  \ell_n=\onehalf\,(-\L^{n-1}+\L^n) \,. \end{array}\ee
At odd level, all orbits (\wrt the full set of simple
currents, which is generated by \js\ for odd $n$, and by \js\ and \jv\
for even $n$) consist of four fields. Each such orbit of integrable
highest weights contains precisely
one representative that satisfies $\L^0\geq\L^1$ and
$\L^{n-1}-\L^n\in2\zet$, implying that $\eli(\L)\in\zet$ and
$k>\ell_1\geq\ell_2\geq\ldots\geq\ell_n$. From now on we restrict
our attention to this particular representative. Thus the numbers
  \be  \tli(\L):=\eli(\L+\rho)=\eli(\L)+n-i  \ee
satisfy $0<\tli(\L)<k+n-1$ for $i=1,2,...\,,n-1$, and $|\tilde\ell_n|<k$.
As it turns out, a special role is played by those orbits for which
$\ell_n=0$; we will refer to such orbits as {\em \sps\/}.
Analogously, orbits that are transformed into each other
upon changing the sign of $\ell_n\ (\neq 0)$ are called
`spinor-conjugate' to each other.

We define now a map $\tau$ between the orbits of \dnk\ and the
tensor orbits of \bke\ as follows. To an orbit of \dnk\ with
representative $\Lambda$ we associate the subset \ml\
of $M=\{1,2,\ldots,k+n\}$ by
  \be  \ml:=\{\,|\tli(\L)|+1\mid i=1,2,...\,,n\} \,. \labl{+1}
Then the (tensor) weight $\tau(\L)$ of \bke\ is defined by the requirement that
the set \mtl \,(with the connection between \L\ and \ml\ for \bke\
defined in the same way as for \bkn\ in section \ref{secbodd}) is given by
  \be  \{\Tli\}\equiv\mtl:=\{k+n+1-l\mid l\in M\setminus\ml\} \,. \labl{taud}
Note that we have chosen our conventions for \dnk\ (in particular the
constant term `+1' in \erf{+1}) in such a manner that the prescription
\erf{taud} is formally the same as \erf{taut} in section \ref{secbodd}.
Furthermore, $\tau(\L)$ is a fixed point iff $k+n\in\mtl$, i.e.\ iff
$1\not\in\ml$, i.e.\ iff \L\ is not \sps. Note also that this map
is {\em not\/} one-to-one on the orbits. Rather, non-\sps\ \dnk\-weights
which transform into each other upon interchanging $\ell^{n-1}$ and
$\ell^n$ get mapped on the same weight of \bke.
(As we will see later on, this is precisely the behavior we need in
\cts\ in order to implement the fixed point resolution.)

We now consider the behavior of the modular matrices $T$ and $S$ under
the map $\tau$.  For the sum of \cdim s one finds
  \be  \begin{array}{ll}  \hl+\htl &=\; [ \sum_{j\in\mL}j^2-(\rho,\rho)
  +\sum_{j\in\mtL}j^2-(\rhot,\rhot)] \,/ [4(k+n-\onehalf)]  \\ {}\\[-2.5 mm]
  &=\;\frac14\,k(k+2n+1)-\frac12\sum_{j\in M\setminus\mL}j \,,
  \end{array}  \ee
which is always a half integer.
The Weyl group of \dnk\ corresponds in the orthonormal basis to
permutations and to even numbers of sign changes of the components,
so that the \kpf\ for the \smat\ leads to
  \be  \sll=(-1)^{n(n-1)/2} 2^{n/2-2} (k+n-\onehalf )^{-n/2} \,
  [\detij\mijpll+\ii^n\,\detij\mijmll] \,, \labl{sdll}
where
  \be  \begin{array}{l} \mijpll:=\cos \mbox{\large[} \dfrac{2\pi\,\eli(\L+
  \rho)\, \ell_j(\Lambda'+ \rho)}{2k+2n-1} \mbox{\large]} , \\ {}\\[-2 mm]
  \mijmll:=\sin \mbox{\large[} \dfrac{2\pi\,\eli(\L + \rho)\,
  \ell_j(\Lambda' +\rho )}{2k+2n-1} \mbox{\large]} \,. \end{array} \ee
Note that $\mijmll=0$ whenever \L\ or $\Lambda'$ are \sps.
For later convenience we
denote by \sllp\ the numbers obtained from \erf{sdll} when
replacing \mijmll\ by zero, i.e.\
  \be  \sllp=(-1)^{n(n-1)/2} 2^{n/2-2} (k+n-\onehalf)^{-n/2} \,
  {\rm det}^{}_{i\in M_\Lambda,j\in M_{\Lambda'}}
  \cos \mbox{\large[} \frac{\pi\,(i-1)(j-1)} {k+n-\onehalf}
  \mbox{\large]}  \,. \labl{sd}

The \smat\ of \bke\ can be calculated analogously as described in
the previous section for \bkn. The result is
  \be  \begin{array}{l}  \stll=(-1)^{k(k-1)/2} 2^{k/2-1} (k+n-\onehalf)
  ^{-k/2} \, \\[1.2 mm] \hsp{28}  \cdot
  {\rm det}^{}_{i\in M\setminus M_\Lambda,j\in M\setminus M_{\Lambda'}}
  \sin \mbox{\large[} {\displaystyle\frac{\pi\,(k+n+\onehalf-i)
  (k+n+\onehalf-j)}{k+n-\onehalf}} \mbox{\large]}  \,. \end{array} \labl{sb}
Combining \erf{sd} with \erf{sb},
we can use the Jacobi-theorem together with the identity
$\sin[\pi\,(k+n+\onehalf-i)(k+n+\onehalf-j)/
(k+n-\onehalf)]=(-1)^{i+j+k+n+1}\cos[\pi\,(i-1)(j-1)/(k+n-\onehalf)]$
to obtain again a relation like \erf{sst}, namely
  \be  \sllp=(-1)^{\Sigma_I+\Sigma_J} \alpha\,(\beta\gamma\delta)^{-1}
  \stll.  \ee
The parameters
are this time calculated as \cite{mnrs}
  \be  \begin{array}{l}  \alpha=(-1)^{k(k-1)/2}(-1)^{k(n+k+1)+\Sigma_I
  +\Sigma_J} 2^{k/2-1}(k+n-\onehalf)^{-k/2}, \\[1.2 mm]
  \beta=(-1)^{n(n-1)/2} 2^{n/2-2} (k+n-\onehalf)^{-n/2},
  \\[1.2 mm] \gamma= 2 (-1)^{(k+n)(k+n-1)/2} ((k+n-\onehalf)/2)^{(k+n)/2},
  \\[1.2 mm] \delta= 2^{-\spS(\Lambda)-\spS(\Lambda')}\,(2/(k+n-\onehalf))^k
  \,,   \end{array}\ee
where
  \be  \spS(\L):=\left\{ \begin{array}{ll} 0 & \mbox{if \L\ is \sps},
  \\[1.2 mm] 1 & {\rm else}.  \end{array}\right. \ee
This leads to
  \be  \sllp= 2^{\spS(\Lambda)+\spS(\Lambda')} \stll.  \labl Q
(When interpreting this equation as a relation between simple current
orbits, one must take the specific representative of the orbit of the $D$
type \wzwt\ described above. Otherwise \erf Q gets modified by a phase.
However, as only tensors of the $B$ type \wzwt\
are involved, the phase does not depend on the representative of
the orbits of the $B$ theory.)
Recalling that $\mijmll=0$, i.e.\ $\sll=\sllp$, if \L\ or $\Lambda'$
are \sps, this means in more detail that
  \be  \stll=\left\{ \begin{array}{ll}  \sll & {\rm for} \ \L \mbox{ and }
  \Lambda' \mbox{ \sps}, \\[1.5 mm] 2\, \sll & {\rm for} \ \L \mbox{ \sps, }
  \Lambda' \mbox{ non-\sps,} \\[.5 mm] & \hsp{62}
  \mbox{or vice versa}, \\[1.5 mm]
  4\,\sllp & {\rm for} \ \L \mbox{ and } \Lambda' \mbox{ non-\sps}.
  \end{array}\right.\ee

\section{$C$ type WZW theories}

When considering $C$ type WZW theories, we are in a more convenient
position than previously. Namely, one can construct a map $\tau$
between individual fields, and not just between simple current
orbits. In the notation of \cite{mnrs}, our map $\tau$ is the
composition of the maps `$\rho$' of section 2 and `tilde' of section
1 of \cite{mnrs}.

We consider again  the components of \L\ in an orthogonal basis of the
weight space. However, for convenience we multiply the components
of the ortho{\em normal\/} basis by a factor $\sqrt2$, because we
then have to deal with integral coefficients only. The components of a
weight \L\ in this non-normalized basis read
  $ \eli(\L)=\sum_{j=i}^{n}\L^j.$
Again we add to these numbers the components of the Weyl vector, i.e.\
define
  \be  \tli(\L):=\eli(\Lambda+\rho)
  =\sum_{j=i}^{n}\L^j+n+1-i.  \labl{53}
This time the integrability condition (\ref{int}) implies, for $(C_n)_k$,
that
  \be  k+n \geq \ele>\ldots>\eli>\ell_{i+1} > \ldots>\ell_n\geq 1 \,.
  \labl{17}
Thus we can describe every weight \L\ uniquely by a set of $n$ positive
integers \tli, $i=1,2,...\,,n$,
subject to $\tli>\tilde\ell_j$ for $i<j$ as well as $\tle\leq k+n$,
that is, by a subset \ml\ of cardinality $n$ of the set
$M=\{1,2,\ldots,k+n\}.$
Given such a subset \ml, we define $\tau(\L)$ through the complementary set
$\{\Tli\}\equiv\mtl:=M\setminus\ml$,
where again the numbers \Tli\ are to be ordered according to
$\Tli>\tilde\ell_j^{(\tau)}$ for $i<j$.
Since this subset of $M$ again satisfies $\Tle\leq k+n$,
and is of cardinality $k$, it describes precisely one
integrable highest weight $\tl$ of \ckn. (In terms of Young tableaux,
the map corresponds to forming the complement \wrt the
rectangular Young tableau $Y(k\la n)$, followed by reflection at an
axis perpendicular to the main diagonal.)

As in the previous sections, it is straightforward to calculate
the quantity $\hl+\htl$. Taking care of the extra factor
$\onehalf$ in the scalar product that is caused by our normalization of
the \eli, one obtains
  \be  \begin{array}{ll}  \hl+\htl &=\; [\onehalf \sum_{j\in\mL}j^2
  -(\rho,\rho) +\onehalf \sum_{j\in\mtL}j^2-(\rhot,\rhot)] \,/ [2(k+n+1)]
  \\ {}\\[-2.5 mm]
  &=\; [ \onehalf \sum_{j=1}^{k+n}j^2-\frac1{12}\,(2 n^3+3n^2+n+2 k^3
  +3k^2+k)] \,/ [2(k+n+1)]  \\ {}\\[-2.5 mm]
  &=\; \frac14\,kn  \,,  \end{array}\labl{518}
where $\rho$ and \rhot\ denote the Weyl vectors of $C_n$ and $C_k$,
respectively.

Proceeding to the modular matrix $S$, we note that
the Weyl group $W$ of $C_n$ acts in the orthogonal basis by permutations
and arbitrary sign changes, implying that
  \be  \sumW{\rm sign}\,(w)\,\exp \mbox{\large[} \frac{\pi\ii}{k+n}\,
  (w(\L+\rho),\Lambda'+\rho) \mbox{\large]} = (2\ii)^n\, \detij
  \mijll  \ee
with
  \be  \mijll:=\sin \mbox{\large[} \frac{\pi\,\eli(\L + \rho )\,
  \ell_j(\Lambda' + \rho )}{k+n+1} \mbox{\large]} \,. \ee
Thus the \kpf\ for the \smat\ yields
  \be  \sll=(-1)^{n(n-1)/2} 2^{n/2} (k+n+1)^{-n/2} \,\detij\mijll,
  \quad \ee
and similarly,
  \be  \stll=(-1)^{k(k-1)/2} 2^{k/2} (k+n+1)^{-k/2} \,\detij\mijtll. \ee
Now \mijll\ can be viewed as a $n\times n$ sub-matrix, and \mijtll\ as a
$k \times k$ submatrix, of the $(k+n)\times (k+n)$ matrix
 $\Om_{ij}:= \sin[(\pi\,i j/(k+n+1)]$,
$i,j\in\{1,2...\,,k+n\}$, such that the two submatrices together
cover each value of $i$ and $j$ precisely once.
As a consequence, the Jacobi-theorem is again applicable, leading
to the relation \erf{sst} between \sll\ and \stll.
The numbers $\alpha,\,\beta,\,\gamma,\,\delta$ in that relation
are this time found to be
  \be  \begin{array}{l}  \alpha=(-1)^{n(n-1)/2}\,2^{n/2} (k+n+1)^{-n/2},
           \qquad        \beta=(-1)^{k(k-1)/2}  2^{k/2} (k+n+1)^{-k/2},
  \\[1.9 mm]  \gamma=(-1)^{(k+n)(k+n-1)/2} ((k+n+1)/2)^{(k+n)/2},
  \qquad      \delta=(2/(k+n+1))^n \,.   \end{array}\ee
When inserting this into \erf{sst}, we make use of the identities
$\Sigma_I= n(n+1)/2+r(\L)$ and $\Sigma_{\Ibar}= k(k+1)/2 + r(\L ')$,
where
  \be  r(\L) := \sum_{i=1}^{n} \eli(\L) \,, \labl{rL}
which is modulo 2 the conjugacy class of the $C_n$-weight \L\
(also, $r$ equals the number of boxes in the Young tableau $Y(\L)$
that is associated to \L). One then obtains
  \be  \sll=(-1)^{r(\Lambda) + r(\Lambda') + k n }\, \stll . \labl{526}

\sect{\nn coset models of type $B$; odd values of rank and level}
\label{sBB} \subsection{The map $\Tau$}

We are now going to describe a one-to-one map $\Tau$ between the
primary fields of the \nn superconformal
\cts\ \Bnk\ and \Bkn. We will show that this map leaves the modular
matrices $S$ and $T$ invariant, and, moreover, provides a one-to-one
map between chiral primary fields. Correspondingly we consider
the two \cts\ as isomorphic \cfts\ and write
  \be  \Bnk \, \stackrel\Tau\cong \, \Bkn \,. \labl{BB}
This is in contrast to the level-rank duality of the underlying
\wzwts\ which is far from providing an isomorphism of conformal
field theories.

To start, let us mention two simple necessary requirements for such
an identification to exist. First, from table \ref{nn} we read off
that the Virasoro central charge of \Bnk\ is
$c_{2n+1,2k+1}^{}=\frac{3(2k+1)(2n+1)}{2(k+n+1)}$,
which is invariant under exchanging $n$ and $k$. It was precisely this
observation \cite{kasu2} that led to the idea of level-rank
duality of these theories. Second, we see that the two theories possess
the same number of (Virasoro and \uE) primary fields. Namely,
for the \ct\ \Bnk\ the number of primaries can be expressed as
  \be  \begin{array}{l} \nu^{BB}_{2n+1,2k+1}= N^D_{2n+1,1} \,N^1_
  {8(k+n+1)} \, \llb \frac1{16}\,[ N^B_{n+1,2k+1}\,N^B_{n,2k+3}-
  F^B_{n+1,2k+1}\,F^B_{n,2k+3}] \\ {}\\[-2.5 mm] \hsp{76}
  +2\cdot\frac14\, F^B_{n+1,2k+1}\,F^B_{n,2k+3} \lrb \end{array}\labl{nBB}
in terms of the numbers $N^B_{m,K}$ of primary fields and $F^B_{m,K}$
of fixed points of the $B$ type \wzwts. Here
the first two factors come from $D_{2n+1}$ at level one and from the
\uE\ theory, respectively. The numbers in the bracket refer to the
theories $B_{n+1}$ at level $2k+1$ and $B_n$ at level $2k+3$;
the term in square brackets corresponds
to the orbits of length four, with the factor $\frac{1}{16}$ taking
care of the selection rule and the identification of order four
(one quarter of the possible combinations of quantum numbers of the
individual theories gets projected out, and each identification
orbit has four members), and the second term
corresponds to the fixed points, the factor of 2 being due to the
resolution procedure (for the fixed points, the factor of $\frac1{16}$
gets replaced by $\frac14$ because none of the fixed points is
projected out by the selection rule encoded in $J_{(1)}$).
Inserting $N^D_{d,1}=4$ and $N^1_{\cal N}={\cal N}$
as well as the formul\ae\ \erf{nB} and \erf{fB} for
$N^B_{m,K}$ and $F^B_{m,K}$, \erf{nBB} becomes
  \be  \nu^{BB}_{2n+1,2k+1}= 2\,\llb 4n+4k+3-\frac{2kn}{k+n+1}\lrb
  \bin{k+n+1}k \bin{k+n+1}n.  \ee
Obviously, for \Bkn\ one obtains the same number of primaries.

After these preliminaries, we now present the map $\Tau$ alluded
to above. Suppose we are given a specific representative
$(\Lambda,{\rm x}\csp\lambda,Q)$
of a field $\Phi$ as described in \erf P;
then we map the simple current orbits of $\Lambda$ and $\lambda$ on
their images under the map $\tau$ that was introduced in section
\ref{secbodd}. Thus
  \be  \Tau(\Phi) \heq (\tau(\lambda),\xtau\csp\tau(\L),\qtau) \,,  \labl=
with $\tau$ as defined in \erf{taus} and \erf{taut}, and with \xtau\
and \qtau\ to be specified below. Now the objects on the right hand side
of \erf= are just representatives of primary fields, and not yet the
primary fields themselves. In particular, the quantities \xtau\ and
\qtau\ are to be considered as orbits, and only after fixing
representatives of the orbits of $\tau(\lambda)$ and $\tau(\L)$,
they are fixed as well so that \xtau\ becomes an element of
$\{$0,\,v,\,s,\,c$\}$ and \qtau\ an integer between 0 and $\cal N$.
To describe the physical fields, we have
to implement the identification currents. According to
table \ref{gid}, in the present case there are two independent
identification currents $J_{(1)}$ and $J_{(2)}$. As $J_{(1)}=(J,1\,/\,J,0)$
acts trivially on the $D_d$ and \uE\ parts, it is convenient
to first restrict the attention to $J_{(1)}$-orbits, and implement
$J_{(2)}$ later on. Provided that no fixed points are present,
 \futnote{Note that in order to have a fixed point of the coset theory,
we must have a fixed point in all \wzwts\ that make up the coset.}
 for fixed choice of \xtau\ and \qtau\
we have to deal with a total of four representatives of two $J_{(1)}$-orbits.

Now observe that, owing to the selection rule implemented by $J_{(1)}$,
the conjugacy classes of \L\ and $\lambda$ coincide, so that
we only need to consider combinations of tensors with tensors, or of
spinors with spinors.
In the case of tensors of both $ (B_{n+1})_{2k+1}$ and $(B_n)_{2k+3}$,
fixed points do not occur. Further, modulo \zet, the conformal dimensions of
the two $J_{(1)}$-orbits differ by \onehalf, precisely as the
conformal dimensions of the corresponding fields of \Bnk.
To start with the definition of $\Tau$, we now simply choose the
$J_{(1)}$-orbit that has conformal weight equal to $\Delta_\Lambda-
\Delta_\lambda$ modulo \zet. Due
to the identification current $J_{(2)}$, this choice actually does
not constitute any loss of generality (but it simplifies some formul\ae\
further on). Namely,
each of the $J_{(1)}$-orbits $\cal O$ lies on the same orbit \wrt $J_{(2)}$
as another $J_{(1)}$-orbit whose values of \xtau\ and \qtau\ differ from
those of $\cal O$ in such a manner that the values of $\Delta_\Lambda
-\Delta_\lambda$ differ by $\onehalf\bmod\zet$.

For spinors, both $J_{(1)}$-orbits in question
have identical conformal weight. The freedom to choose one
of the orbits turns out to be closely connected with the issue
of fixed point resolution. Namely, the property of $\tau$
to map \wzw fixed points on WZW-orbits of length two and vice versa,
translates into the following property of $\Tau$: any `unresolved
fixed point' is mapped on two distinct fields, and vice versa, such that the
non-fixed points of one theory precisely describe the resolved
fixed points of the other theory. In case that just one of the orbits
in either the `numerator' or the `denominator' of the \ct\ is a fixed
point, we have exactly the reversed situation in the dual theory.

Having fixed the $B$ parts of the theory, we extend
the definition of $\Tau$ to the \uE\ and $D_d$ parts
by the following definitions:
the $D_d$ part remains unchanged, i.e. $\xtau=$\,x, while the \uE-charge
is transformed according to
  \be \qtau = -Q + \left \{ \begin{array}{ll}
  QL \qquad & \mfor \quad \cl = 0 \quad \mand \quad  \xns \,, \\ [1.1 mm]
  Q(2n+1)L \qquad & \mfor\quad\cl=0 \quad \mand \quad \xr \,,  \\ [1.1 mm]
  Q(2k+1)L \qquad & \mfor\quad\cl=1 \quad \mand \quad \xns \,, \\ [1.1 mm]
  (2n-2k-Q)L \qquad & \mfor \quad \cl = 1 \quad \mand \quad  \xr \,.
  \end{array} \right. \labl-
Here, for convenience, we use the abbreviation
  \be  L = 2(k+n+1) \,,\ee
and all \uE-charges are understood modulo ${\cal N} = 4L$.
(Thus $L$ is one quarter of the \uE-charge of the primary field
that extends the chiral algebra of the \uE\ theory, and hence the
appearance of this number in \erf- is quite as natural.)

The definition of $\Tau$ is not yet complete, of course, as we still
have to make precise its meaning when acting on, or mapping to,
resolved fixed points.
Nevertheless already at this stage we can verify that
$\Tau$ as defined above satisfies the following properties: \\[2 mm]
1.  The result is independent of the particular choice of the
    representative of the original field $\Phi$.
 \futnote{Also, applying the analogous prescription $\Tau_\Tau$ to the
  transformed field $\Tau(\Phi)$ brings us back to the field $\Phi$
  of the original theory, thus justifying the name {\em du}ality.} \\[2 mm]
2.  The conformal weights $\Delta$ of fields related by $\Tau$
    are equal modulo \zet, which implies that the modular $T$-matrices of
    the two theories coincide. This is in fact already the maximal
    information about conformal dimensions that
    we could hope to prove in the general case, because for primary
    fields of a \ct\ (other
    than \rgs s of an \nn theory) it is very hard to compute the integer
    part of the conformal weight. \\[2 mm]
3.  The superconformal \uE-charges coincide modulo 2 (again, except for
    \rgs s it is
    hard to show that the charges coincide exactly).\\[2 mm]
    Actually, the two last-mentioned properties (together with a
    prescribed choice of the orbits of $\tau(\L)$ and $\tau(\lambda)$,
    such as the one discussed above) already specify uniquely
    \xtau\ and \qtau\ for fields that are not fixed points. Thus our
    choice $\xtau=\,$x and \qtau\ as in \erf- is the only possibility
    that allows for $\Tau$ to possess the required properties. \\[2mm]
4.  The elements of the modular $S$-matrices corresponding to
    non-fixed points coincide after properly
    taking into account the field identification.
    As we will show in the next subsection, the same is true for fixed points;
    it follows that both theories possess the same fusion rules, and,
    together with the first observation, that their characters realize
    isomorphic representations of SL$(2,\zet)$.
    If the $B$ weights of one field are tensors and those of the other
    field are spinors, \erf{fc} implies that the corresponding
    \smat\ element of the full theory is simply the product of the respective
    WZW \smat\ elements if the spinors are fixed points, and twice this
    product if the spinors are not fixed points. For the dual theory,
    the corresponding factor of two is provided by our map $\Tau$ through
    the factor $\sqrt2$ that appears (both for the `numerator' and the
    `denominator' of the coset theory) in the transformation
    \erf{sts} of \smat\ elements of the $B$ type \wzwts\ under $\tau$.
    \\[2mm]
5.  $\Tau$ maps the unique \rgs\ $\Phi_R^{\rm max}$ with highest
    superconformal charge $q=\frac c6$ of one theory to the
    corresponding \rgs\ of the dual theory. (This check is particularly
    important, as this field is the simple current that
    generates spectral flow.) Namely, for this field there
    is a standard representative \cite{fuSc} $\Phi_R^{\rm max}\heq
    (0,{\rm s}\csp\rhog-\rhoh)$, and $\Tau$ maps this particular
    representative to the analogous representative of the highest
    \rgs\ of the dual theory.  \\[1mm]

\subsection{Fixed points}

In order to prove that these statements pertain to the full \cts,
 \futnote{Recall that only after fixed point resolution, we are
allowed to interpret the object \cosetK{\tilde g}{\tilde h}
as a genuine \cft.}
 we now come to the more detailed description of the action of
$\Tau$ on fixed points, as promised.
(The fixed point resolution will be interesting also
from a different point of view, see the remarks after \erf x below.)
As it turns out, this is a somewhat subtle issue.
We will first deal with the case when an unresolved fixed point is
mapped on a pair of non-fixed points.
In fact, we have so far only specified on what pair of fields a fixed point
gets mapped, and noticed that the number of the fields is the right one.
But each unresolved fixed point gives rise to two distinct
physical fields, and so we have to describe
which of the resolved fixed points is mapped to which field.
To settle this question, it is not sufficient to look at the fractional part
of the conformal dimensions
$\Delta$ and superconfomal charges $q$, because for the two resolved
fixed points the conformal dimensions and superconformal charges must
coincide modulo \zet\ and 2\zet, respectively.
Thus we have to resort to the modular matrix $S$.

In order to simplify notation, we first look at those parts
of the theory which behave non-trivially under the identification current
that has fixed points, which is
$J_{(1)}=(J,1\,/\,J,0)$. In other words, we restrict our attention
to the theory
$(B_{n+1})_{2k+1}[(B_n)_{2k + 3}]^*$, where we use the symbol `*' to
indicate that the complex conjugates of the modular $S$- and $T$-matrices
are to be considered (compare the remarks after \erf{:}).
As has been shown in \cite{scya6}, the matrices $\Gamma_{(\cdot)}$
appearing in \erf{fc} and
in the factorization \erf{fac} are given, up to certain phases,
by the $S$-matrices of the \wzwts\ $(C_n )_k$ and $(C_{n-1})_{k+1}$.
We denote these phases, to be determined below, by $\omega_n$ and
$\omega_{n-1}$, respectively.

In terms of the components \tli, the relation between fixed points
and the corresponding fields of the fixed point theory is given by
  \be \tilde\ell_i^{(C)} = \tilde\ell_{i+1}^{(B)}   \labl{cb}
for $i=1,2,...\,,n$. In other words, for the $S$-matrices
the resolution of fixed points amounts
to simply deleting the row and the column with $i = k+n+1$ of the matrix
\Om\ as defined in \erf{Om}. But it was precisely this row that
made the \smat\ elements vanish if fixed points were involved.
Now once more we can use the Jacobi-theorem for the $(k+n) \times
(k+n)$ matrix $M_{ij} = \sin[(\pi ij)/(k+n+1)]$
to relate the \smat\ of the fixed point resolution to the
\smat\ of the images of the fixed points. We find that
  \be \tilde S_{\Lambda \Lambda'} \tilde S_{\lambda \lambda'} =
  \eps \omega_{n-1}\omega_n\, (-1)^{\Sigma_{\Lambda\Lambda'}+\Sigma
   _{\lambda\lambda'}+1 }
  S_{\tau(\Lambda) \tau(\Lambda')}  S_{\tau(\lambda) \tau( \lambda')} \,.
  \label{fix} \labl{traB}
Here $\tilde S$ denotes the \smat\ of the fixed point resolution, while
  \be  \Sigma_{\Lambda\Lambda'}^{}=\sum_{i\in\mL}i +\sum_{i\in
  M_{\Lambda'}}i \,, \labl s
and $\Sigma_{\lambda\lambda'}$ is the sum of the analogous numbers
for the theory in the `denominator.' Further, $\eps\equiv\eps_{\Lambda
\lambda\,\Lambda'\lambda'}^{}\in\{1,-1\}$ depends
on the particular action of $\Tau$ on resolved fixed points. Namely,
the left hand side of \erf{traB} is to be multiplied with the matrix
$P=\onehalf\biN{\,\ 1\;-1}{-1\,\ \;1}$. On the right hand side of \erf{traB}
this is reflected by the fact that the subscripts actually do not
refer to an orbit, but to a specific representative; the sign of the
right hand side changes when one changes from one representative to the other
representative of the orbit. The two representatives, which will be
denoted by \tgll\ and \tkll,
can be described as follows. For any orbit $\{\phi_\Lambda,J\phi_\Lambda\}$
of a $B$ type \wzwt\ denote by $\Lambda_<$ the representative
with smaller values of \tle, and by $\Lambda_>$ the other one; then
$\tgll:=(\tau(\L)_<,\tau(\lambda)_<)\cong (\tau(\L)_>,\tau(\lambda)_>)$,
while $\tkll:=(\tau(\L)_<,\tau(\lambda)_>)\cong (\tau(\L)_>,\tau
(\lambda)_<)$, with the two equivalent states mapped onto one another
by the action of the identification current $J_{(1)}$.
Now the value of $\eps$ depends on whether the first of the resolved
fixed points is mapped to \tgll\ and the second to \tkll, or the
other way round. As we will see, a consistent prescription for this
choice can be given for which $\eps$ precisely cancels the further
possible signs in \erf{traB}.

To compute the phases $\omega_n$ and $\omega_{n-1}$,
we first note that, given a representation
  $(ST)^3 = S^2$,\, $S^4 =\one$
of SL$(2,\zet)$, the only rescalings of $S$
and $T$ which again lead to a representation of SL$(2,\zet)$ are
  \be T \mapsto \eE^{\pi\ii m/6}\, T \,,\qquad \quad
  S \mapsto \eE^{-\pi\ii m/2}\, S  .  \labl;
We can determine the integer $m$ in the first of these rescalings from the
global shift in the conformal dimensions that is present in the
fixed point theories as compared to the $C$ type \wzwts. In the
case of our interest we have for $(B_{n+1})_{2k+1}$ the shift
$\Delta_{(B)}-\Delta_{(C)} =(6k+2n+3)/24$, and analogously
for $(B_n)_{2k+3}$. Subtracting the two shifts, one finds $m =-2$. With
\erf;, this implies that for the resolution one should take minus the product
of the $S$-matrices of the $C$ type theories rather than simply their
product. In other words, $\omega_{n-1}\omega_n=-1$, and hence
\erf{traB} reduces to
  \be \tilde S_{\Lambda \Lambda'} \tilde S_{\lambda \lambda'} = \eps
  (-1)^{\Sigma_{\Lambda\Lambda'}+\Sigma_{\lambda\lambda'}}
   S_{\tau(\Lambda) \tau(\Lambda')} S_{\tau(\lambda)\tau(\lambda')}
  \,. \labl(

To complete the construction of $\Tau$, we first investigate the
restrictions that are
obtained from requiring that the \smat\ is left invariant.
Let us choose an arbitrary fixed point $\Phi_f\heq(\L,\lambda)$ to
start with, and denote
the resolved fixed points by $\Phi_{f_\pm}$, as in \erf{pm}. We can now
map $\Phi_{f_+}$ either to \tgll\ or to \tkll\ (and, correspondingly,
$\Phi_{f_-}$ to \tkll\ and to \tgll, respectively).
After fixing this choice, the requirement that the \smat\ should be
invariant already fixes $\Tau(\Phi_{f'})$ for any fixed point $\Phi_{f'}$
uniquely. Namely, assume that the first possibility,
$\Phi_{f_+}\mapsto\tgll$, is chosen; then we have to map
$\Phi_{f'_+}\mapsto\tglp$, $\Phi_{f'_-}\mapsto\tklp$ if the number
$\Sigma_{ff'}\equiv\Sigma_{\Lambda\Lambda'}+\Sigma_{\lambda\lambda'}$
computed according to \erf s is even,
while if $\Sigma_{ff'}$ is odd, the map must be
$\Phi_{f'_+}\mapsto\tklp$, $\Phi_{f'_-}\mapsto\tglp$.
With this prescription, one obtains $\eps^{}_{ff'}=(-1)^{\Sigma_{ff'}}$, and
hence \erf( reduces to the desired equality
  \be \tilde S^{}_{\Lambda \Lambda'} \tilde S^{}_{\lambda \lambda'}\,P_{ij}^{}
  = \Llb S^{}_{\tau(\Lambda) \tau(\Lambda')} S^{}_{\tau(\lambda)
  \tau(\lambda')} \Lrb_{\Tau(i)\,\Tau(j)} \,, \ee
where on the left hand side $i,j\in\{+,-\}$,
while on the right hand side $\Tau(i),\Tau(j)\in\{<,>\}$.
This not only works for any fixed choice of $f'$, but also for
all \smat\ elements $S_{f'f''}$, because $\Sigma_{f'f''}=\Sigma_{ff'}
+\Sigma_{ff''}$. The latter identity also implies that the choice
of reference fixed point $\Phi_f$ is immaterial.

As long as we only take care of the \smat , the alternative to choose
$\Phi_{f_+}\mapsto\tgll$ or $\Phi_{f_+}\mapsto\tkll$ means that
there are two different allowed mappings on the fixed points.
But according to \erf{pm} the characters of $\Phi_{f_+}$
and $\Phi_{f_-}$ are different; $\Phi_{f_+}$ has more states
with minimal conformal weight. Therefore by looking at the characters
one can remove the ambiguity in the definition of $\Tau$.
However, since this reasoning can be applied to any fixed point,
it has also to be checked whether the constraints obtained from
different fixed points are compatible.
In practice, this is quite difficult to check, as it requires a
detailed analysis of the characters. But there is a rather general
argument that the consistency conditions coming from the characters
are compatible with those originating from the \smat.
Namely, defining for any fixed point $f$ the function
  \be  {\cal X}^{}_{\tau(f)}:=\chi^{}_{\tau(f)_>}-\chi^{}_{\tau(f)_<} ,\labl x
it is easy to verify that the functions $\cal X$ transform under
the modular group
exactly like the character modifications $\ckri\equiv(\chi^{}_{f_+}-
\chi^{}_{f_-})/2v$. In itself, this does not yet imply that ${\cal X}^{}
_{\tau(f)}$ and $\ckri$ are necessarily equal, but the fact that the
result holds for an infinite series is a rather strong hint that they
indeed coincide. (Note that it directly follows from \erf{pm} that
only $\cal X$ as defined in \erf x, and not $-{\cal X}$ can be a
sensible character;
thus there is in particular no sign ambiguity in defining ${\cal X}$.)

In principle, we should perform the same kind of reasoning as above
also for resolved fixed points that
occur as the images of non-fixed points. However, due to the duality property
of the map $\Tau$ the arguments needed for this analysis closely parallel
the arguments given above, so that we refrain from repeating them here.

At this point it is worth to recall that there does not exist a
general proof that the fixed points of a \ct\ can be resolved in
a unique way \cite{scya6}. In the present case, the manner in which
the resolution procedure described in \cite{sche3}
fits to the duality map $\Tau$ is however so non-trivial, that
it is hard to imagine that there could exist another
prescription for the resolution that would be compatible with
duality as well. Note that
the \epop s of the theories considered here should
obey level-rank duality for {\em any\/} possible resolution,
because according to
quite general arguments \cite{sche3,fuSc} the \epop\ of an \nn \ct\
does not depend on the details of the resolution procedure.

\subsection{\rgs s}

Finally we turn our attention to the chiral ring of the theories.
According to the formula \erf{levw}, the number of representatives of \rgs s
with a fixed $(B_{n+1})_{2k+1}$ weight is given by the relative size
  \be \frac{|W(g)|}{|W(h)|} = \frac{2^{n+1}\,(n+1)!}{2^n\,n!}
  = 2\,(n+1)    \ee
of the Weyl groups.
After implementing the resolution of fixed points, one finds that
the dimension of the ring is indeed invariant under the exchange
of $n$ and $k$; this is a direct consequence of the much stronger
result \cite{sche3} that the (ordinary, and also even
the extended) \pop s of the theories coincide.

Our goal is now to show that the map $\Tau$ defined above maps
every \rgs\ to a \rgs\ of the dual theory with identical
superconformal charge. To do so, we first note
that the relation \erf{ltw} between \L\ and $\lambda$ can be
reformulated in terms of the sets \ml\
and $M_\lambda$, and of the charge $Q$, as follows. Take a highest
$g$-weight $\Lambda$ described by the set \ml, and consider it as
ordered with respect to the magnitude of the elements. The action of
any Weyl group element $w$ is then
to permute the elements of \ml\ and to multiply them with a sign:
the $2(n+1)$ special elements of the classes of ${W(g)}/{W(h)}$
that appear in \erf{ltw} are characterized by the property that they
choose among the $n+1$ elements of \ml\ a
particular element \tli\ which gets placed before all the other
elements, and change its sign or not, leaving all other signs
unchanged. We will denote such a Weyl group element that
maps the $i$th basis vector $e_i$ of the orthonormal basis on $\pm e_1$
and respects the ordering of all other basis vectors by $w_i^{(\pm)}$.
By inserting the explicit form of the roots $\alpha$ in the orthonormal
basis into \erf{HUmp}, it is straightforward to calculate the length of the
elements $w_i^{(\pm)}$. We find
  \be l(w_i^{(+)} ) = i - 1 \qquad \mbox{and} \qquad
  l(w_i^{(-)} ) = 2 (n+1) - i   \,,     \labl{blen}
where $n+1$ is the rank of the algebra. This result reflects the
linear structure of the associated Hasse diagram of the embedding
$B_{n}\hookrightarrow B_{n+1}$ \cite{ekmy}.

For the \rgs\ defined by acting with $w_i^{(\pm)}$ on $\Lambda$, the
\uE-charge $Q$ is given by
$\pm 2\tli$ for spinors and $\pm(2\tli - 1)$ for tensors.
Opposite sign choices correspond to choosing charge-conjugate \rgs s.
As a consequence, the map $\Tau$ automatically respects the charge
conjugation properties of the \rgs s, and hence is compatible with
the conjugation isomorphisms of the chiral rings of the theories.
As mentioned in the introduction, this compatibility must in fact hold
on rather general grounds.

Next we remark that not all representatives of a \rgs\ are of the
form \erf{ltw} (recall that \erf{ltw} is a formula for
representatives, and not for physical fields). To be able to employ
the relation \erf{ltw}, we therefore pick a specific
representative of any combination of simple current orbits of
weights \L\ and $\lambda$ that describes a \rgs.
After applying the map $\Tau$
in the form \erf=, \erf- to this specific representative
of a \rgs\ $\Phi_R$, we obtain a specific representative
of the primary field $\Tau(\Phi_R)$ of the dual theory. What we have
to show is that $\Tau(\Phi_R)$ is again a \rgs, and we will do this
by employing the formula \erf{ltw}. Of course, generically the
particular representative of $\Tau(\Phi_R)$ with which we are
dealing in the first place cannot be expected to be of the form
\erf{ltw}. As we will see, it is indeed sometimes not of this form,
but as was shown in \cite{levw} there is always at least one representative
of the \rgs\ fulfilling \erf{ltw}.

Suppose, to start with, that $\Lambda$ and $\lambda$ are both spinor weights,
and that the \rgs\ is given by the Weyl group element $w_i^{(+)}$ acting
on $\Lambda$. Recalling that the index $i$ of $w_i^{(+)}$ refers to the
fact that $\ml \setminus M_\lambda= \{ \tli \}$, and observing that
via the map $\tau$ on the \wzwts, i.e.\
upon forming the complement relative to $\{ 1,2,...\,, k+n+1 \}$,
this is transformed to the relation
$M_{\tau(\lambda)} \setminus \mtL = \{ \tli \}$, we learn that there
exists a Weyl group element \wtau\ of the dual theory that relates
$\tau(\lambda)$ and $\tau(\L)$ in the correct manner and is given by
one of the two elements $w_{i^{}_\Tau}^{(\pm)}$, with $i_\Tau^{}$
determined by the requirement $\tilde\ell^{(\tau)}_{i^{}_\Tau}=\tli.$
To decide which of these two  elements is the correct one, we observe
that owing to the latter relation \qtau\ must be equal either to $Q$ or
to $-Q$; from \erf- (together with the explicit form of the
identifiation currents) it follows that in fact $\qtau= -Q$.
In summary, using the sets $M_{\tau(\lambda)}$ and $M_{\tau(\Lambda)}$,
and the sign of \qtau\ relative to the sign of $Q$, we
fix a unique Weyl group element \wtau\ of $W(B_{k+1})$;
in fact, a more detailed analysis shows that $i^{}_\tau=k+n-Q/2-i+3$, i.e.
$w_{\Tau} = w^{(-)}_{k+n-Q/2 -i + 3}$.
To verify that this Weyl group element indeed provides us with a \rgs,
the only thing that we still have to do is
to check that it yields the proper $D_d$ part.
 \futnote{In some cases we also must show that the correct
$J_{(1)}$-orbit out of two possibilities is chosen. This happens
when an `unresolved fixed point' gets resolved into two
fields whose conformal weights differ by an integer. The discussion of
fixed points in the previous subsection shows that indeed the right
orbit is chosen.}
 While in the foregoing discussion we fixed the representative \wrt $J_{(2)}$
by $\xtau=$\,x, the present choice of representative for the charge $\qtau$
implies that \xtau\ must be given by
  \be  \xtau=(\jv)^{n - k - Q/2} {\rm x} \,. \labl X
Now the formul\ae\ \erf{blen} for the length of Weyl group elements tell
us that
  \be l(w) - l(w_\Tau) = n - k - Q/2 \,, \labl{ldb}
and hence, recalling that the sign of $w$ is equal to $(-1)^{l(w)}$,
  \be \sign(w) \: \sign(\wtau) = (-1)^{l(w) + l(w^{}_\Tau)}
  = (-1)^{k+n+Q/2}. \ee
In view of \erf c, this shows that \erf X is indeed fulfilled.
Furthermore, plugging \erf{ldb} into the formula \erf{Len} for the
superconformal charge of \rgs s, it follows that $\Phi_R$ and
$\Tau(\Phi_R)$ have the same superconformal charge
(exactly, and not just modulo 2).

The reasoning above applies also to the case $w = w_i^{(-)}$,
as the two cases are clearly dual to each other.
If both $\Lambda$ and $\lambda$ are tensor weights, the situation is slightly
more complicated. This is because \tli\ gets mapped under $\tau$
to $\tilde\ell^{(\tau)}_{i^{}_\Tau}=\frac L2+1-\tli$.
If the \rgs\ is defined by $w = w_i^{(+)}$, this shows that $Q= 2\tli -1$
should be mapped on $\qtau= L-Q $, implying that $w_\Tau$ involves no
minus sign. While in the foregoing discussion we always chose the
representative of the field by requiring that $\Delta_\Lambda-\Delta_\lambda$
should be an integer, we now have to
fix the representative by requiring that $\qtau=-Q+L$,
which, owing to the second identification current $J_{(2)}$,
is always possible. This choice of representative leads to
  \be  \xtau = (\jv)^{n-(Q -1)/2}{\rm x}\,.   \labl J
Again, a Weyl group element \wtau\ for the dual theory is completely
fixed, and can be shown to be given by $\wtau = w^{(+)}_{(Q+1)/2 -n+i-1} $ .
It follows that $l(w)-l(\wtau)= n - \frac{Q+1}2 + 1$, so that
  \be \sign(w)\: \sign(\wtau) = (-1)^{n+(1-Q)/2}\,,   \ee
implying that the correct mapping \erf J of the $D_d$-weights is reproduced,
and also that the superconformal charge is left invariant.
It is also clear that we have chosen the right $J_{(1)}$-orbit,
because $\Delta$ is conserved modulo \zet\ under $\Tau$ and
because the relevant different $J_{(1)}$-orbits differ in their
conformal weight by \onehalf\ modulo \zet.

For $w = w_i^{(-)}$, the discussion must be slightly
changed. This time $Q = -(2\tli -1)$ is mapped on $\qtau = -L -Q$, i.e.\ we
have to choose a different representative, leading to
x$' =(\jv)^{n+2k+(Q+1)/2}$ x. Explicit calculation shows that
$\wtau = w_{i-n -(Q+1)/2}^{(-)}$, leading to $l(w) - l(\wtau)=
n-2k-\frac{Q+1}2$, which gives the right transformation of the $D_d$ part
and implies identity of superconformal charges.

Thus we have proven that $\Tau$ always maps \rgs s to \rgs s with identical
superconformal charge.

\sect{Type $B$ coset models with level and rank not congruent modulo 2}

In the same spirit as before, we can deal
with the other level-rank dualities mentioned in the introduction.
As the discussion often closely parallels the one of the previous
section, we will usually be rather brief and shall only mention some new
features. In the present section we use the map $\tau$ for
$B$ type algebras at even level to relate the \ct\
$(B, 2k+1, 2n)$ with the $D$ type modular invariant to $(B, 2n, 2k+1)$
with the diagonal modular invariant, i.e.\
  \be (B, 2k+1, 2n)_{\mid D} \, \stackrel\Tau\cong \, (B,2n, 2k+1) \,.
  \labl{DD}
According to subsection 2.2, taking the $D$-invariant amounts to
incorporating the integer spin
simple current $J_{(3)}:=(J,1\,/\,1,0)$ into the chiral algebra. This
introduces further fixed points which can have order 2 or 4 and which
have to be resolved, but it also has the crucial
advantage that it leaves us with tensors of
the $B$ algebras only, so that the map $\tau$ constructed
in section 4 is applicable.

The choice of the $J_{(1)}$-orbits is now
immaterial. This is because the presence of $J_{(3)}$ implies that
$\tkll\cong\tgll$, so that any pair of tensor orbits of the $B$ type \wzwts,
combined with a $D_d$-weight and a \uE-charge, corresponds to
a single physical field. However, we still have to take into
account the additional identification current $J_{(2)}\star J_{(3)}=
(1, \jv\,/\, 1, \pm 2L)$, where $L:=2k+2n+1$.

Again the general form of the map $\Tau$ is given by \erf= (recall that
on the right hand side of \erf= only a representative of $\Tau(\Phi)$
is given).
Starting from a fixed representative of a field or a fixed point $\Phi$
of the $B$ type \ct\ at even level and odd rank, we obtain all
representatives of $\Tau(\Phi)$ by using the map $\tau$ and the
identification currents of the \ct\ at even rank and odd level.
Moreover, with the help of the identification currents
we can also fix uniquely a representative of $\Tau(\Phi)$ for which
$\tau(\L)$ and $\tau(\lambda)$ are tensors and which has the
same conformal weight as the chosen representative of $\Phi$.
Note that fixed points are mapped
on a spinor-conjugate pair of orbits, which reflects
the resolution of fixed points. In particular fixed
points of order two and of order four are mapped on two and four
fields, respectively.

One can now show again that there is a unique mapping $\Tau$ that
preserves both the superconformal charge $q$ modulo 2 and the
conformal dimension $\Delta$ modulo integers; it is given by
  \be  \qtau = \left \{ \begin{array}{lll}  -Q + QL&\mfor &  \xns\,, \\
  -Q +(2k+1)Q L   &\mfor &  \xr\,,  \end{array} \right. \ee
and
  \be \xtau= \left \{ \begin{array}{lll} (\jv)^{Q/2} {\rm x} &\mfor& \xns\,, \\
  (\jv)^{-k+(Q-1)/2} {\rm x} &\mfor &  \xr\,.    \end{array} \right. \ee
To check this, one has to make use of the fact that the representatives
of the orbits of the $D$ type \wzwts\ that were chosen above always
have vanishing monodromy charge relative to $(\js,1\,/\,\js,0)$.

Of course, again $\Tau$ must be complemented by a prescription on
the fixed points. This time the fixed point theory is not a \wzwt;
rather, it is closely related to certain \cfts, denoted by the
symbol $\cal B$,
that were described in \cite{scya6}. In fact, the existence of the
map $\Tau$ suggests that the \smat\ and characters of the $\cal B$
theories are related to a $D$ type \wzwt,
and it should be interesting to explore the level-rank duality further
to gain deeper insight in the structure of these peculiar \cfts.
Finally, it is again possible
to prove that the modular $S$-matrices are identical and that \rgs s are
mapped on \rgs s with equal superconformal charge.

\sect{$BB$ versus $CC$ theories}

In this section we present the isomorphism
  \be (BB, n+2, 1) \stackrel\Tau\cong (CC, 2, 2n+1) \,.  \labl{BC}
To relate the non-\hsc s $(BB, n+2, 1)$ and $(CC, 2, 2n+1)$ we first
notice the isomorphism $C_2 \cong B_2$ of simple Lie \alg s. This allows to
make use once again of the map $\tau$ of section 3 to relate the
$(B_n)_5$ theory appearing in $(BB, n+2, 1)$ with the $(B_2)_{2n+1}
\cong(C_2)_{2n+1}$ part of $(CC,2,2n+1)$. The $(B_{n+2})_1$ part, on the other
hand, is comparatively easy to deal with, because it has only three
integrable highest weights, and because the identification current
$J_{(1)}$ strongly restricts their combination with weights of the
other parts. Namely, $(B_n)_5$-weights that are tensors must be combined with
either the tensor weight $\L=0$ or the tensor weight $\L=\Lambda_{(1)}$
of $(B_{n+2})_1$, while spinors are to be combined with the spinor
weight $\Lambda_{(n+2)}$ of $(B_{n+2})_1$; furthermore, $J_{(1)}$
introduces an additional
identification, implying that in the case of tensors we can characterize
the $B$ part completely by a $(B_n)_5$-weight and by the difference
$\Delta_\Lambda-\Delta_\lambda$ of the conformal dimensions.
Also, by using the identification current $J_{(1)}$ of the $CC$ models,
we can choose without loss of generality for a fixed representative of
$\Phi$ the representative of the $C_2$-orbit in such a way that it has
conformal dimension $\Delta_\Lambda-\Delta_\lambda$
modulo integers. For spinor fixed points we have again an
ambiguity which is connected to the issue of fixed point resolution.

This time, the mapping $\tau$ has to be complemented not only by
a mapping on the $D_d$ and \uE\ parts, but also on the $(A_1)_{2n+2}$ part
of the theory. Thus
  \be  \begin{array}{l}  \Phi \heq (\L,{\rm x}\csp\lambda,\mu,Q) \,, \\[2.2 mm]
  \Tau(\Phi) \heq (\tau(\lambda),\xtau\csp\mu_\Tau^{},\qtau) \,,
  \end{array}\ee
where $\mu$ and $\mu_\Tau$ are $A_1$-weights (recall that $C_1\cong A_1$).
It is easy to see that equality of the superconformal charges modulo 2
is equivalent to the relation $\xtau=(\jv)^{Q} {\rm x}$.
In fact one can show again that there is a unique mapping that preserves
the fractional part of $\Delta$, as well as $q$ modulo 2. Namely,
choosing the weights of the $B$ parts in the manner decribed above,
for tensors in the $B$ parts one needs
  \be \qtau= \left \{ \begin{array}{lll}  -Q + QL&\mfor &  \xns \,, \\
                   -Q + (Q+1) L    &\mfor &  \xr  \end{array}
 \right. \ee
with $L=2n+4$, while for spinor weights in the $B$ parts we must set
  \be \qtau= \left \{ \begin{array}{lll}  -Q + L&\mfor &  \xns \,, \\
        -Q  &\mfor &  \xr \,.  \end{array} \right. \hsp{10.9} \ee
The corresponding prescription for the weight $\mu$ of $(A_1)_{2m+2}$
is, independent of the value of x,
  \be \mu^{}_\Tau= \left \{ \begin{array}{ll} J^\mu\mu &
  \mfor \ c_\Lambda^{}=c_\lambda^{}=0 \,, \\[1 mm]
  \mu &\mfor\ c_\Lambda^{}=c_\lambda^{}=1 \,. \end{array}
  \right.\hsp{16.7} \ee

Fixed points have to be dealt with more carefully again. Using general
simple current arguments, it is easy to see that the \smat\ element
between a fixed point and {\em any} other spinor has to vanish.
At first sight, this might seem inconsistent, because the \smat\ element
between two non-fixed point spinors of $(B_n)_5$ does not vanish in general,
whereas both are mapped on fixed points \wrt $J_{(1)}$ of the $C_2$-theory,
and the \smat\ of the image vanishes. However, spinors of $(B_n)_5$ are
always combined with the spinor weight $\Lambda_{(n+2)}$ of
$(B_{m+2})_1$; now $S_{\Lambda_{(n+2)}\Lambda_{(n+2)}}^{}$
vanishes, and hence the same is true for the corresponding \smat\
element of the \ct.

We can use the Jacobi-theorem to relate the \smat\ arising in the
resolution of the fixed points to the \smat\ of the $CC$ theory.
The resolution is this time accomplished
by mapping the fixed point on an orbit of length two.
Calculation shows that the product of the $S$-matrix elements of $A_1$, $D_d$,
and \uE\ differs from the corresponding \smat-element of the $CC$ \ct\
by a factor of $\eps(-1)^{P + Q}$, where $P$ and $Q$ are the \uE-charges
of the $BB$ theory, and where the sign $\eps$
depends on the specific action of $\Tau$ on fixed points analogously
as discussed after \erf s. In a similiar
manner as we dealt with the factor $(-1)^{\Sigma}$ in section 6, it
can be shown that the action of $\Tau$ can be chosen in such a way that
$\eps(-1)^{P + Q}$ is the correct sign for obtaining equality of the full
$S$-matrices. A parallel argument also shows that this definition of
$\Tau$ reproduces the correct identification between the characters of the
resolved fixed points and those of the corresponding fields of the $CC$
theory. Let us also mention that the factors stemming from the \smat\ of
$(B_{m+2})_1$ precisely compensate the different size of the identification
group in the case of non-fixed points; for fixed points they assure, together
with the factors of $\sqrt2$ appearing in \erf{sts},
the equality of the $S$-matrices.

It is by now not too difficult to verify that the mapping $\Tau$
fulfills the same properties as in the cases treated in the previous
sections. Besides preserving $q$ and $\Delta$ as
well as the modular $S$-matrix, we see that $\Tau$ maps again the \rgs s
with highest superconformal charge onto each other, proving again the
isomorphisms of the spectral flows. It is also possible to check
that \rgs s are mapped on \rgs s with the same superconformal charge
quite in the same way we did before.
Owing to the presence of the additional $A_1$-subalgebra,
the arguments are, however, slightly more complicated, and we
refrain from presenting the technical details here.

\sect{Duality in the $CC$ series}

Here we construct a map $\Tau$ between the \nn superconformal coset
models \Cnk\ and \Ckn, which as in the previously discussed cases leaves
$S$ and $T$ invariant and identifies the rings of chiral primary fields,
  \be  \Cnk \, \stackrel\Tau\cong \, \Ckn \,. \labl{CC}
The definition of the map $\Tau$ will be such that
  \be  \Tau((\L,{\rm x}\csp\lambda,Q)) = (\tau(\lambda),\xtau\csp\tau(\L),
  \qtau) \,. \labl|
This is formally very similar to the analogous definition \erf= in
section 6, but its contents is quite different. Namely,
this time the underlying map $\tau$ of the $C$ type \wzwts\
was defined on representatives of simple current orbits rather
than on the orbits themselves (see section 5). Correspondingly, \erf|
is a map between representatives as well, and hence we will
have to check that the relevant quantities of the \cts\ do not
depend on the choices of representatives. Therefore we will be
a bit more explicit than in the two previous sections.

We begin again by checking the conformal central charge and the
number of primaries. According to table \ref{nn},
the Virasoro charge of \Cnk\ is equal to $-3+6n(k+1)/(k+n+1)$,
and hence is invariant under exchanging $n\leftrightarrow k+1$.
The number of primaries of the \cnk\ \wzwt\ is $N^C_{n,k}=
\biN{n+k}{k}$. Furthermore, the \ct\ does not have any fixed points,
and hence the number of primary fields of \Cnk\ is
  \be  \nu^{CC}_{n,k}= \frac14\,N^C_{n,k}\,N^D_{2n-1,1}\,N^C_{n-1,k+1}
  \,N^1_{2(k+n+1)}= 2(k+n+1)\, \bin{k+n}n \,\bin{k+n}{n-1} , \labl F
where the first factor $\frac14$ takes care of the selection rule and
the identification of order two. Obviously,
the number of primaries of the \Ckn\ theory is given by \erf F, too.

Next we present the map $\Tau$.
In \erf|, $\tau$ is to be taken as the map defined after \erf{17}, and
\xtau\ and \qtau\ are defined by
  \be  \xtau= \left\{ \begin{array}{lll}
  (\jv)^{k+n+Q+1} {\rm x} & {\rm for} & {\rm x}\in\{{\rm s,c}\} \,, \\[1 mm]
  (\jv)^{k+1-Q} {\rm x} & {\rm for} & {\rm x}\in\{{\rm v,0}\} \,,
  \end{array}\right. \hsp{4.5} \labl{cvz}
and
  \be  \qtau= \left\{ \begin{array}{lll}
  -Q & {\rm for} & {\rm x}\in\{{\rm s,c}\} \,, \\[1 mm]
  -Q+k+n+1 & {\rm for} & {\rm x}\in\{{\rm v,0}\} \,.  \end{array}\right.\ee
Note that already in terms of representatives, the map $\Tau$ squares
to the identity, $\Tau_\Tau^{}\circ\Tau={\sl id}$. Also,
combining the expression \erf h for the conformal dimension of $\Phi$ with
the result \erf{518} for the conformal dimensions of the $C$ type
\wzwts, one can again show that $\Tau$ is the only map that
preserves $q$ modulo 2 and the fractional part of the conformal
weight $\Delta$, as well as the \smat . To check the last-mentioned property,
it is important make use of the selection rules encoded in the
identification current $J_{(1)}$.

As already emphasized, the map $\Tau$ must provide
a mapping between fields rather than only
a mapping between formal combinations of weights of the underlying Lie
algebras. The following remarks show that the mapping is indeed well
defined on physical fields. \\[2 mm]
 1. The map \erf| is consistent with the selection rules, i.e.\ it maps
allowed fields to allowed fields. Note that the dependence of \qtau\
on x is necessary to fulfill the selection rule encoded in $J_{(1)}$,
(explicitly, the selection rule
reads $r( \Lambda ) + r( \lambda ) + n\sigma+ Q \equiv 0 \bmod 2$,
where $r(\L)$ is the number defined in \erf{rL}, which modulo 2
is equal to the conjugacy class of \L, and where
$\sigma$ is 0 in the \NS sector and 1 in the Ramond sector). \\[2 mm]
 2. Identification currents are mapped onto identification currents:
 \futnote{This does not furnish a group isomorphism between
the groups that describe the fusion rules of the identification
currents. Since these groups are isomorphic to $\zet_ 2$, such an
isomorphism would necessarily be trivial.}
  \be \begin{array}{l}
  (0,0\csp0,0) \,\stackrel\Tau\longleftrightarrow\, ((n-1)\Lambda_{(k+1)}^{},
    (\jv)^{k+1}_{}\csp n\Lambda_{(k)}^{},\pm(k+n+1)) \,, \\[2.5 mm]
  (k\Lambda_{(n)}^{}, (\jv)^{n}_{}\csp (k+1)\Lambda_{(n-1)}^{},\pm(k+n+1))
  \,\stackrel\Tau\longleftrightarrow\, (0,0\csp0,0) \,.  \end{array}\ee
Computation shows that the products of $S$-matrices of the
respective \wzwts\ coincide
(one has to make use once again of the selection rules, which
imply cancellation of the factors $(-1)^{r(\Lambda)}$ that are present
in equation \erf{526}). This implies that in fact the two
representatives of one physical field
$\Phi$ are mapped on the representatives of the corresponding physical
field $\Tau(\Phi)$ of the dual theory, or, in other words, that we can
interpret $\Tau$ also as a mapping of physical fields. \\[2 mm]
 3. The two representatives of the \rgs\ with highest \uE-charge
get exchanged:
  \be \begin{array}{l} (0,{\rm s}\csp0,n)
  \,\stackrel\Tau\longleftrightarrow\, ((n-1)\Lambda_{(k+1)}^{},
    (\jv)^{k+1}_{}{\rm s}\csp n\Lambda_{(k)}^{},-n) \,, \\[2.1 mm]
  (k\Lambda_{(n)}^{}, (\jv)^{n}_{}{\rm s}\csp (k+1)\Lambda_{(n-1)}^{},-k-1)
  \,\stackrel\Tau\longleftrightarrow\, (0,{\rm s}\csp0,k+1) \,.  \end{array}\ee
In other words, in terms of fields we have proven compatibility of
the map $\Tau$ with spectral flow. \\[1 mm]

To show that $\Tau$ maps \rgs s on \rgs s, again we first check the
dimension of the chiral ring. We have to use the formula \erf{levw} with
$N=N_{n,k}^C$, $|Z|=|Z(C_n)|=2$, and
  \be \frac{|W(g)|}{|W(h)|} = \frac{2^n\,n!}{2^{n-1} (n-1)!} = 2n \,.\ee
Thus $\mu^{CC}_{n,k}= n\,N^C_{n,k}=(n+k)!/((n-1)!\,k!)$,
which is invariant under $n\leftrightarrow k+1$.
Of course, this also follows from the fact, observed in
\cite{fuSc}, that the (ordinary and extended) \pop s of the
theories \Cnk\ and \Ckn\ are identical.

To analyse the \rgs s in more detail, first recall that in the
orthogonal basis the action of the Weyl group is given
by permuting the components and multiplying them with a sign,
and has thus the same structure as in the case of $B$ type Lie algebras.
This allows us to use the same notation for Weyl group elements as in
section 6. Furthermore, the roots of $B$ type and $C$ type algebras
differ only by normalization factors, and these are irrelevant for
the determination of the length of Weyl group elements.
As a consequence, the formul\ae\ \erf{blen} are valid
for $C$ type Lie algebras, too (and the Hasse diagram of the embedding
$C_{n-1}\hookrightarrow C_n$ is again linear \cite{fuSc}).
Correspondingly, the reasoning below will
be very similar to the one of section 6.
The relation \erf{ltw} between the weights \L\ and $\tilde\lambda$
implies that in terms of the numbers \tli\ introduced in \erf{53},
the $C_{n-1}$-weight $\lambda$ of a \rgs\ $\Phi_R$
is related to the $C_n$-weight \L\ by
  \be \tli (\lambda) = \tilde\ell_{i+1} (w(\Lambda)) \,,    \ee
and also
  \be |Q| = \tle (w(\Lambda))      \ee
for some Weyl group element $w$.
When we characterize $\Lambda$ and $\lambda$ by the sets
\ml\ and $M_\lambda$, this translates into
  \be M_\lambda = \ml \setminus \{ \lto \} , \ee
where $\lto = \pm Q$ is an arbitrary element of \ml\ (recall that
$\lto > 0$). Again the freedom
in the choice of the sign of $Q$ reflects the invariance of the set of
\rgs s under charge conjugation.
An analogous description applies to the image $\Tau(\Phi_R)$ of the
\rgs. Now $\Tau$ fixes uniquely the transformation of all weights, and
  \be M_{\tau(\lambda)} = M \setminus M_{\lambda} = ( M \setminus
  M_{\Lambda} ) \cup \{\lto \} =M_{\tau(\Lambda)} \cup \{\lto\}  ,  \ee
so that $\tau(\Lambda)$ and \qtau\ are related to $\tau(\lambda)$
by the formula \erf{ltw} with a suitably chosen Weyl group element
$w^{}_\Tau$.

To verify that $\Tau(\Phi_R)$ is again a \rgs, it is
now sufficient to check that $\Tau$ gives the correct weight in the
$D_d$ part of the theory. The Weyl group elements $w$ and $w^{}_\Tau$
are uniquely fixed by the weights $\Lambda$ and $\tilde\lambda$,
respectively by their images under $\tau$;
for $w = w_i^{(+)}$, the Weyl group element $w_\Tau$ is given by
$w_{k+n -Q-i+2}^{(-)}$ , which implies that
$   l(w) - l(w_\Tau) = n - k - 1 -Q $. From this equation we can derive
not only the equality of superconformal charges, but also the behavior on
the $D_d$ part; we have
 \be \sign(w)\:\sign(\wtau)=(-1)^{l(w)+l(w^{}_\Tau)}= (-1)^{k+n+1+Q}     , \ee
which reproduces the prescription given in \erf{cvz}.
This shows that $\Tau$ maps \rgs s on \rgs s, as claimed, and thus
completes our arguments that the map $\Tau$ fulfills the requirements for
the isomorphism \erf{CC} of \cfts, analogously as for the other
isomorphisms of \erf1.

\vskip 9 mm \noindent
{\bf Acknowledgements.}~~We are grateful to M.\ Kreuzer,
W.\ Lerche, A.N.\ Schellekens, and M.G.\ Schmidt for helpful comments,
and to E.J.\ Mlawer, S.G.\ Naculich, H.A.\ Riggs, and H.J.\ Schnitzer
for pointing out that, contrary to what we claimed in an earlier
version of this paper, our results for WZW level-rank dualities
are in perfect agreement with those of \cite{mnrs}.

 \end{document}